\documentclass[12pt,titlepage]{article}
\usepackage[utf8]{inputenc}

\usepackage{amsmath}
\usepackage{amsfonts}
\usepackage{amssymb}
\usepackage{graphicx}
\usepackage{cite}
\usepackage{endfloat}
\usepackage{mychicago}
\usepackage{subfigure}
\usepackage{genetics_manu_style}
\usepackage{authblk}
\usepackage{lineno}

  \def\P{\mathbb{P}}
  \def\E{\mathbb{E}}
  \def\I{\mathbb{I}}

\title{Demographic inference using genetic data from a single individual: separating population size variation from population structure}

\author[*]{Mazet Olivier}
\author[*]{Rodríguez Valcarce Willy}
\author[$\S$,$\dagger$,$\ddagger$]{Chikhi Lounès}
\affil[*]{Université de Toulouse, Institut National des Sciences Appliquées, Institut de Mathématiques de Toulouse, F-31077 Toulouse, France}
\affil[$\S$]{CNRS, Université Paul Sabatier, ENFA, UMR 5174 EDB (Laboratoire Évolution \& Diversité Biologique), F-31062 Toulouse, France}
\affil[$\dagger$]{Université de Toulouse, UPS, EDB, F-31062 Toulouse, France}
\affil[$\ddagger$]{Instituto Gulbenkian de Ciência, P-2780-156 Oeiras, Portugal}
\renewcommand{\RunningHead}{Structure vs Population Size Change}

\renewcommand{\CorrespondingAddress}{
  CNRS, Université Paul Sabatier, Laboratoire Evolution \& Diversité Biologique, Bâtiment 4R1, 118 route de Narbonne, 31062 Toulouse cedex 9, France.
  \texttt{lounes.chikhi@univ-tlse3.fr} \vfill}

\renewcommand{\CorrespondingAuthor}{Chikhi Lounès}
\renewcommand{\KeyWords}{symmetric island model, population size change, maximum likelihood estimation, demographic history, coalescence time}

\date{}

\begin{document}
\maketitle
%\tableofcontents

%\linenumbers

%////////////////Abstract/////////////////

\begin{abstract}

The rapid development of sequencing technologies represents new opportunities for population genetics research. It is expected that genomic data will increase our ability to reconstruct the history of populations. While this increase in genetic information will likely help biologists and anthropologists to reconstruct the demographic history of populations, it also represents new challenges. Recent work has shown that structured populations generate signals of population size change. As a consequence it is often difficult to determine whether demographic events such as expansions or contractions (bottlenecks) inferred from genetic data are real or due to the fact that populations are structured in nature. Given that few inferential methods allow us to account for that structure, and that genomic data will necessarily increase the precision of parameter estimates, it is important to develop new approaches. In the present study we analyse two demographic models. The first is a model of instantaneous population size change whereas the second is the classical symmetric island model. We (i) re-derive the distribution of coalescence times under the two models for a sample of size two, (ii) use a maximum likelihood approach to estimate the parameters of these models (iii) validate this estimation procedure under a wide array of parameter combinations, (iv) implement and validate a model choice procedure by using a Kolmogorov-Smirnov test. Altogether we show that it is possible to estimate parameters under several models and perform efficient model choice using genetic data from a single diploid individual.

\end{abstract}

%////////////////Abstract/////////////////

%////////////////Introduction/////////////////
\section{Introduction}

The sheer amount of genomic data that is becoming available for many organisms with the rapid development of sequencing technologies represents new opportunities for population genetics research. It is hoped that genomic data will increase our ability to reconstruct the history of populations \cite{li2011inference} and detect, identify and quantify selection \cite{vitti2013detecting}. While this increase in genetic information will likely help biologists and anthropologists to reconstruct the demographic history of populations, it also exposes old challenges in the field of population genetics. In particular, it becomes increasingly necessary to understand how genetic data observed in present-day populations are influenced by a variety of factors such as population size changes, population structure and gene flow \cite{nielsen2009statistical}. Indeed, the use of genomic data does not necessary lead to an improvement of statistical inference. If the model assumed to make statistical inference is fundamentally mis-specified, then increasing the amount of data will lead to increased precision for perhaps misleading if not meaningless parameters and will not reveal new insights  \cite{nielsen2009statistical,chikhi2010confounding,heller2013confounding}.

For instance, several recent studies have shown that the genealogy of genes sampled from a deme in an island model is similar to that of genes sampled from a non structured isolated population submitted to a demographic bottleneck \cite{peter2010distinguishing,chikhi2010confounding,heller2013confounding}. As a consequence, using a model of population size change for a spatially structured population may falsely lead to the inference of major population size changes \cite{stadler2009impact,peter2010distinguishing,chikhi2010confounding,heller2013confounding,paz2013demographic}. Conversely, assuming a structured model to estimate rates of gene flow when a population has been submitted to a population size change, may also generate misleading conclusions, even though the latter case has been much less documented.  More generally, previous studies have shown that spatial processes can mimic selection \cite{currat2006comment}, population size changes \cite{chikhi2010confounding,heller2013confounding} or that changes in gene flow patterns can mimic changes in population size \cite{wakeley1999nonequilibrium,broquet2010genetic}. The fact that such dissimilar processes can generate similar coalescent trees poses exciting challenges \cite{nielsen2009statistical}. One key issue here is that it may be crucial to identify the kind of model (or family of models) that should be used before estimating and interpreting parameters.

One solution to this problem is to identify the "best" model among a set of competing models. This research program has been facilitated by the development of approximate Bayesian computation (ABC) methods \cite{pritchard2000inference,beaumont2002approximate,cornuet2008inferring,beaumont2010approximate}. For instance, using an ABC approach, \citeN{peter2010distinguishing} showed that data sets produced under population structure can be discriminated from those produced under a population size change by using up to two hundred microsatellite loci genotyped for 25 individual. In some cases, relatively few loci may be sufficient to identify the most likely model \cite{sousa2012,peter2010distinguishing}, but in others, tens or hundreds of loci may be necessary \cite{peter2010distinguishing}. ABC approaches are thus potentially very powerful but it may still be important to improve our understanding of the coalescent under structured models.

In the present study we are interested in describing the properties of the coalescent under two models of population size change and population structure, respectively, and in devising a new statistical test and estimation procedures. More specifically we re-derive the full distribution of  $T_2$, the time to the most recent common ancestor for a sample of size two for a model of sudden population size change and for the \textit{n-island} model. We then use a maximum likelihood approach to estimate the parameters of interest for each model (timing and ratio of population size change former and rate of gene flow and deme size for the latter). We develop a statistical test that identifies data sets generated under the two models. Finally, we discuss how these results may apply to genomic data and how they could be extended to real data sets (since $T_2$ is not usually known) and other demographic models. In particular we discuss how our results are relevant in the context of the PSMC (Pairwise Sequentially Markovian Coalescent) method \cite{li2011inference}, which has been now extensively used on genomic data and also uses a sample size of two.

%////////////////Introduction/////////////////

%////////////////Demographic_Models/////////////////
\section{Methods}
\subsection{Demographic models}

\subsubsection{Population size change:}
\begin{figure}
  \centering
  \includegraphics[width=\textwidth]{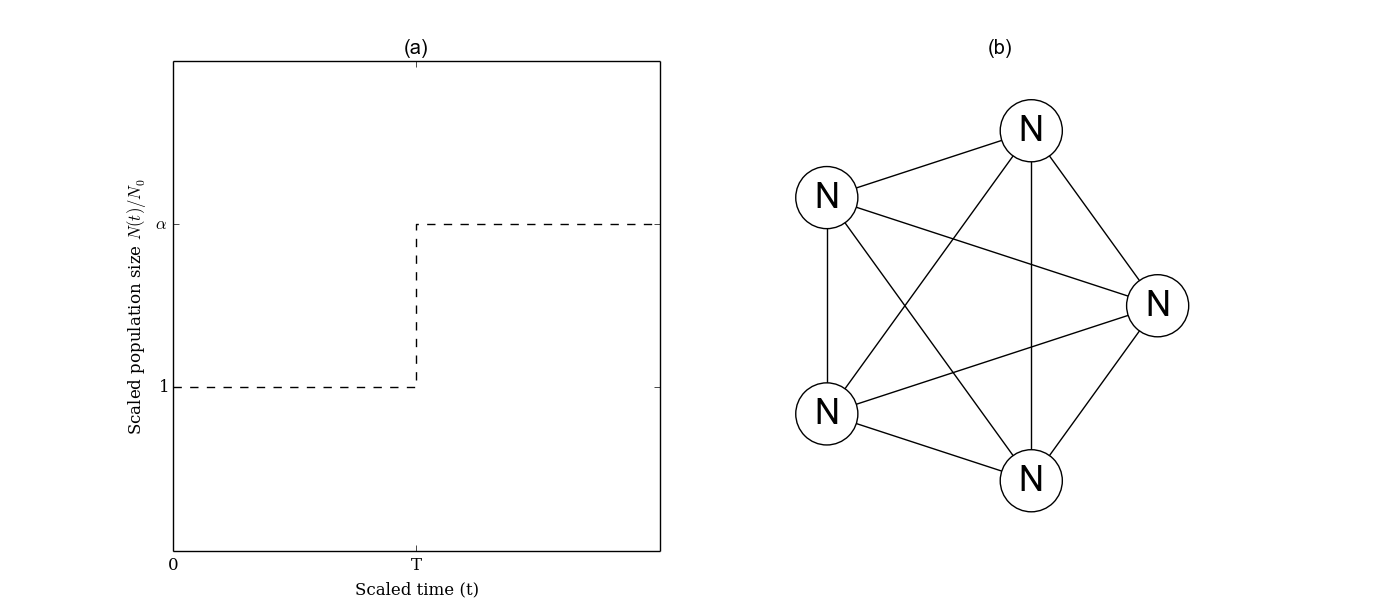} 
  \caption{Demographic models. (a): Single step population size change (SSPSC) model. The x-axis represents $t$, the time to the past in units of generations scaled by the number of genes. At time $t=T$, the population size changes instantaneously from $N_1$ to $N_0$ by a factor $\alpha$. The y-axis represents the population sizes in units of $N_0$ (\textit{i.e.} $N(t)/N(0)$). (b): Structured symmetrical island (StSI) model for $n=5$ islands. Each circle represents a deme of size $N$. All demes are connected to each other by symmetrical gene flow, represented by the edges. In this example the total number of genes is $5N$.}
  \label{fig:Bottleneck_Struct_graphic}

\end{figure}

We consider a simple model of population size change, where $N(t)$ represents the population size ($N$, in units of genes or haploid genomes) as a function of time ($t$) expressed in generations scaled by $N$, the population size, and where $t=0$ is the present, and positive values represent the past (Figure \ref{fig:Bottleneck_Struct_graphic} (a)).  More specifically we assume a sudden change in population size at time $T$ in the past, where $N$ changes instantaneously by a factor $\alpha$. This can be summarized as $N(t)=N(0)=N_0$ for $t\in [0,T[$, $N(t)=N(T)=\alpha N_0$ for $t\in [T,+\infty[$. If $\alpha>1$ the population went through a bottleneck (Figure \ref{fig:Bottleneck_Struct_graphic}) whereas if $\alpha<1$ it expanded. Since $N$ represents the population size in terms of haploid genomes, the number of individuals will therefore be $N/2$ for diploid species. Note also that for a population of constant size the expected coalescence time of two genes is $N$ generations, which therefore corresponds to $t=1$. We call this model the SSPSC, which stands for Single Step Population Size Change.

\subsubsection{Structured population:}
Here we consider the classical symmetric \textit{n-island} model \cite{wright1931evolution}, see Figure \ref{fig:Bottleneck_Struct_graphic} (b), where we have a set of $n$ islands (or demes) of constant size $N$, interconnected by gene flow with a migration rate $m$, where $M=Nm$ is the number of immigrants (genes) in each island every generation. The whole metapopulation size is therefore $nN$ (this is the total number of genes, not the effective size). Again, $N$ is the number of haploid genomes, and $N/2$ the number of diploid individuals. We call this model the StSI, which stands for Structured Symmetrical Island model.

%////////////////Demographic_Models/////////////////

%////////////////Distribution_Moments/////////////////
\subsection{The distribution of coalescence times: qualitative and quantitative analyses}
\paragraph{}
In this section we used previous results \cite{herbots1994stochastic,donnelly1995coalescents} to derive the distribution of coalescent times for the two models of interest. We show that even though they are different, these distributions can be similar under an indefinitely large number of parameter values (Figures \ref{fig:ExpVarCompare} and \ref{fig:DensityCompare}). Moreover we show that even when the distributions are distinguishable, their first moments may not be. In particular, we show that the first two moments (mean and variance) are near identical for a large number of parameter combinations. Before doing that we start by providing a simple intuitive rationale explaining why and how a model of population structure can be mistaken for a model of population size change. This intuitive approach is important because it allows us to understand how the parameters of the two models (($T$, $\alpha$) and ($M$, $n$), respectively) are linked.

\subsubsection{Intuitive and qualitative rationale:}

We start by taking two genes sampled in the present-day population under the Single Step Population Size Change (SSPSC) model. If we assume that $\alpha>1$ (population bottleneck from an ancient population of size $N_1$ to a current population of size $N_0$, with $N_1 = \alpha N_0$) the probability that the two genes coalesce will vary with time as a function of $N_0$, $N_1$ and $T$. If $T$ is very small, then most genes will coalesce at a rate determined by $N_1$, whereas if $T$ is very large the coalescence rate will be mostly determined by $N_0$. If we now take two genes sampled from the same island in the Structured Symmetrical Island (StSI) model, we can also see that their coalescence rate will depend on $N$, the size of the island and on $m$, the migration rate. If $m$ is very low, the coalescence rate should mostly depend on $N$. If $m$ is high, the two genes may see their lineages in different islands before they coalesce. As a consequence the coalescence rate will depend on the whole set of islands and therefore on the product $nN$, where $n$ is the total number of islands.

This intuitive description suggests that there is an intrinsic relationship between $T$ and $1/M$, and between $\alpha$ and $n$. The reason why structured populations exhibit signals of bottlenecks is because in the recent past the coalescence rate depends on the local island size $N$, whereas in a more distant past it depends on $nN$. In other words, it is as if the population size had been reduced by a factor of $n$. As we will see this rationale is only qualitatively correct, but it suggests that if we want to distinguish them it may be necessary to derive the full distribution of the coalescence times under the two models. We shall denote these coalescence times $T_2^{SSPSC}$ and $T_2^{StSI}$, respectively.

\subsubsection{Derivation of the distribution of coalescence times:}

\paragraph{The distribution of $T^{SSPSC}_2$:}
\label{eq:pdf_T2_SSPCS}

The generalisation of the coalescent in populations of variable size was first rigorously treated in \citeN{donnelly1995coalescents}), and is clearly exposed in \citeN{tavare2004part}). 
%If we set to $N(k)$ the population size $k$ generations ago, for any $k$, and if we assume that $\frac{N(k)}{N(0)}$ (i.e. the ratio to present-day population size, $N_0$) converges to a finite number for each $k$ when $N(0)$ goes to infinity, 
If we denote by $\lambda(t)$ the ratio $\frac{N(t)}{N(0)}$ where $t$ is the time scaled by the number of genes (\textit{i.e.} units of coalescence time, corresponding to $\lfloor N(0)t\rfloor$ generations), we can compute the probability density function ($pdf$) $f^{SSPSC}_{T_2}(t)$ of the coalescence time $T^{SSPSC}_2$ of two genes sampled in the present-day population. Indeed, the probability that two genes will coalesce at a time greater than $t$ is

\begin{equation}
\P(T^{SSPSC}_2>t)=e^{-\int_0^t \frac{1}{\lambda(x)}dx}\ ,
\end{equation}
 
where 
$$
\lambda(x)=\I_{[0,T[}(x)+\alpha\I_{[T,+\infty[}(x),
$$
and $\I_{[a,b[}(x)$ is the Kronecker index such that 
$$
\I_{[a,b[}(x) = \begin{cases}
                 1 & \text{ for }x \in [a,b[ \\
                 0 & \text{ otherwise.} 
                \end{cases}
$$

Given that the $pdf$ is 
$$
f^{SSPSC}_{T_2}(t)=(1-\P(T^{SSPSC}_2>t))'
$$
Equation (1) can be rewritten as
$$
\P(T^{SSPSC}_2>t)=e^{-t} \I_{[0, T[} + e^{-T-\frac{1}{\alpha}(t-T)} \I_{[T, +\infty[} .
$$
This leads to the following $pdf$

\begin{equation}
\label{SSPSCT2density}
f^{SSPSC}_{T_2}(t)=e^{-t} \I_{[0, T[}(t) + \frac{1}{\alpha} e^{-T-\frac{1}{\alpha}(t-T)} \I_{[T, +\infty[}(t).
\end{equation}

\paragraph{The distribution of $T^{StSI}_2$:}
\label{eq:pdf_T2_StSI}

Following \citeN{herbots1994stochastic}), an easy way to derive the distribution of the coalescence time $T_2^{StSI}$ of two genes for our structured model, is to compute the probability that two genes are identical by descent when they are sampled from the same or from different populations. These two probabilities are respectively denoted by $p_s(\theta)$ and $p_d(\theta)$, where $\theta=2uN$ is the scaled mutation rate, $u$ being the \textit{per} locus mutation rate.

Indeed, using a classical scaling argument (see for instance \citeN{tavare2004part}, page 34), we can note that 
$$p_s(\theta)=\E(e^{-\theta T_2^{StSI}})$$
In other words $p_s(\theta)$ is the Laplace transform of $T_2^{StSI}$.

We can compute this probability as follows. Taking two genes from the same island and going back in time, there are three events that may occur: a coalescence event (with rate $1$), a mutation event (with rate $\theta$) and a migration event (with rate $M$). Taking now two genes from different islands, they cannot coalesce and therefore only a mutation or a migration event may occur. Migration events can then bring the lineages in the same island with probability $\frac{1}{n-1}$, and in different islands with probability $\frac{n-2}{n-1}$. We thus obtain the following coupled equations:

$$
p_s(\theta)=\frac{1}{1+M+\theta}+\frac{M}{1+M+\theta}p_d(\theta),
$$
and
$$
p_d(\theta)=\frac{M/(n-1)}{M+\theta}p_s(\theta)+\frac{M(n-2)/(n-1)}{M+\theta}p_d(\theta).
$$
By solving them, we obtain
$$
p_s(\theta)=\frac{\theta+\gamma}{D} \text{  and  }
p_d(\theta)=\frac{\gamma}{D}
$$
with 
$$
\gamma=\frac{M}{n-1}
\text{  and  }
D=\theta^2+\theta(1+n\gamma)+\gamma.
$$
We can then obtain the full distribution through the Laplace transform formula, if we note that
$$
p_s(\theta)=\frac{\theta+\gamma}{(\theta+\alpha)(\theta+\beta)}=\frac{a}{\theta+\alpha}+\frac{1-a}{\theta+\beta}
$$
with
$$
a=\frac{\gamma-\alpha}{\beta-\alpha}=\frac{1}{2}+\frac{1+(n-2)\gamma}{2\sqrt{\Delta}},
$$

where

$$
\alpha=\frac{1}{2}\left(1+n\gamma+\sqrt{\Delta}\right)
$$
and
$$
\beta=\frac{1}{2}\left(1+n\gamma-\sqrt{\Delta}\right).
$$
Noting now that for any $\theta$ and any $\alpha$ we have 
$$
\int_0^{+\infty}e^{-\alpha s}e^{-\theta s}\, ds=\frac{1}{\theta+\alpha},
$$
it is straightforward to see that the $pdf$ of $T_2^{StSI}$ is an exponential mixture:

\begin{equation}
\label{StSIT2density}
f^{StSI}_{T_2}(t)=ae^{-\alpha t}+(1-a)e^{-\beta t}.
\end{equation}

\subsubsection{First moments:}

Equations \ref{SSPSCT2density} and \ref{StSIT2density} are different hence showing that it is in principle possible to identify genetic data produced under the two demographic models of interest. The two equations can be used to derive the expectation and variance of the two random variables of interest, $T_2^{SSPSC}$ and $T_2^{StSI}$. Their analytic values can be easily expressed as functions of the model parameters:

\begin{equation*}
   \label{eq:ExpVarianceBottleneck}
  \begin{aligned}
     \E\left(T_2^{SSPSC}\right)= & 1+e^{-T}(\alpha-1), \\ 
     Var\left(T_2^{SSPSC}\right)= & 1+2Te^{-T}(\alpha-1)+2\alpha e^{-T}(\alpha-1)-(\alpha-1)^2e^{-2T}, \\
     \E(T_2^{StSI})= & n, \\
     Var(T_2^{StSI})= & n^2+\frac{2(n-1)^2}{M}.
  \end{aligned}
\end{equation*}

It is interesting to note that the expected time in the StSI model is $n$ and does not depend on the migration rate \cite{durrett2008probability}. The variance is however, and as expected, a function of both $n$ and $M$. For the SSPSC model, the expected coalescence time is a function of both $T$ and $\alpha$. We note that it is close to $1$ when $T$ is very large and to $\alpha$ when $T$ is close to zero. Indeed, when the population size change is very ancient, even if $\alpha$ is very large the expected coalescence time will mostly depend on the present-day population size, $N_0$. Similarly, when $T$ is small it will mostly depend on $N_1$. The relationship that we mentioned above between $n$ and $\alpha$ (and between $M$ and $1/T$) can be seen by noting that when $T$ is close to zero (and $M$ is large), the expectations under the two models are $\alpha$ and $n$, and the variances are $Var\left(T_2^{SSPSC}\right) \approx 1+2\alpha(\alpha-1)-(\alpha-1)^2 = \alpha^{2}$ and $Var(T_2^{StSI}) \approx n^2$. This exemplifies the intuitive rationale presented above. This relationship is approximate and will be explored below, but can be illustrated in more general terms by identifying scenarios with similar moments.

\begin{figure}[h]
  \centering
  \includegraphics[width=\textwidth]{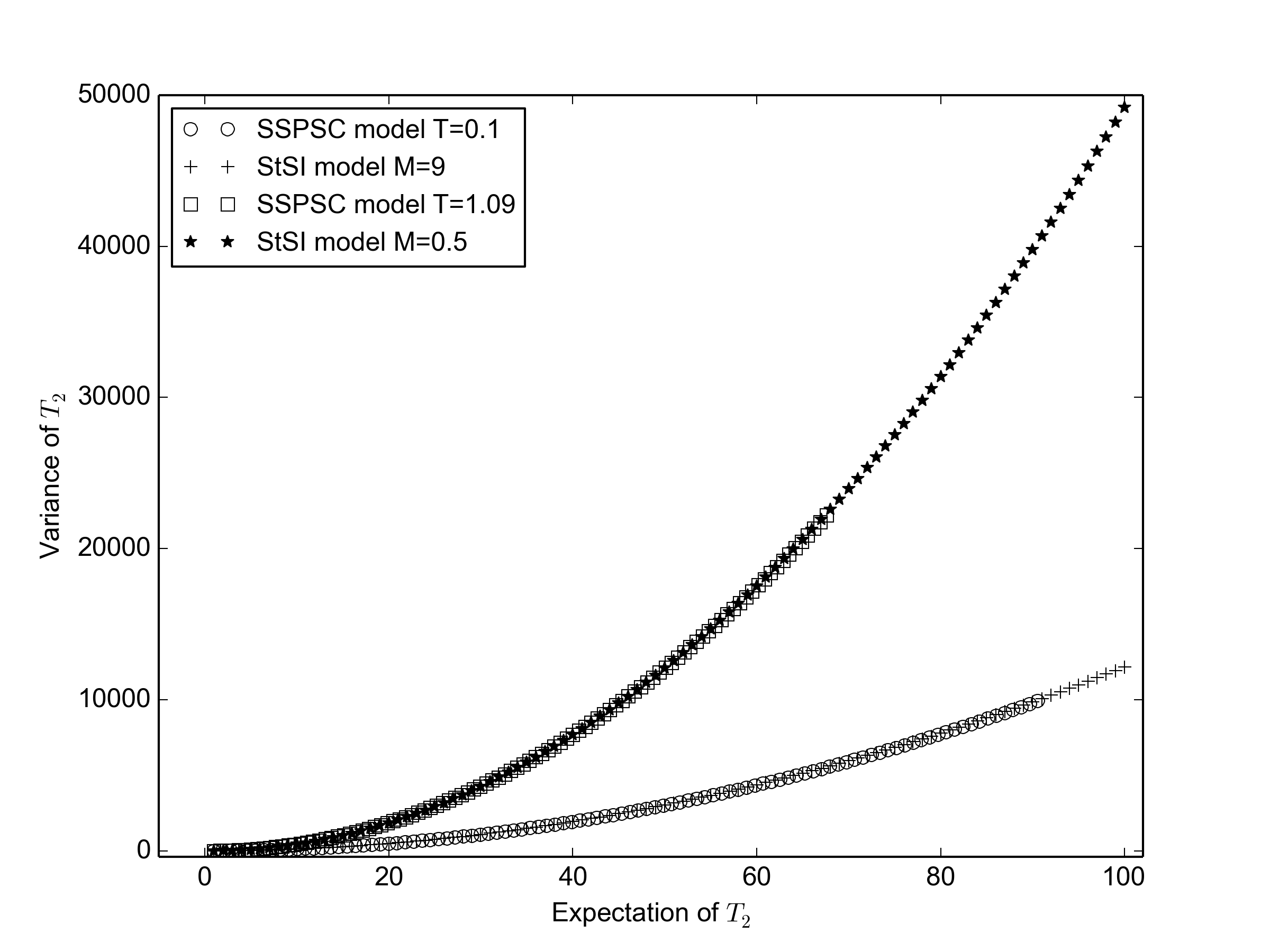}
  \caption{Expected value and Variance of $T_2$ under the SSPSC and StSI models. This figure ilustrates how both models can have the same pair of values $\left(E(T_2), Var(T_2)\right)$ for many sets of parameters. For the SSPSC model the time at which the population size change occured was fixed to $T=0.1$ whereas $\alpha$ varied from $1$ to $100$ in one case, and $T=1.09$, whereas $\alpha$ varied from $1$ to $200$ in the other case. For the StSI model the migration rate was fixed to $M=9$ and $M=0.5$, whereas $n$ varies from $2$ to $100$. }
  \label{fig:ExpVarCompare}
\end{figure}

As figure \ref{fig:ExpVarCompare} shows, the two models provide near-identical pairs of values for $(\E(T_2), Var(T_2))$ for ``well chosen'' parameters $(T,\alpha)$ and $(M,n)$. Here by setting $T$ to $0.1$ (and $M$ to $9$, \textit{i.e.} $1/M \approx 0.11$) whereas $\alpha$ and $n$ were allowed to vary from $1$ to $100$, and from $2$ to $100$, respectively, we see that the two models exhibit very similar behaviours. We also plotted a second example obtained by setting $M$ to $0.5$ and $T$ to $1.09$, and varying $n$ and $\alpha$ as above. These examples ilustrate how $n$ and $\alpha$ (respectively,  $M$ and $1/T$)  are intimately related.

\begin{figure}[h]
  \centering
  \includegraphics[width=\textwidth]{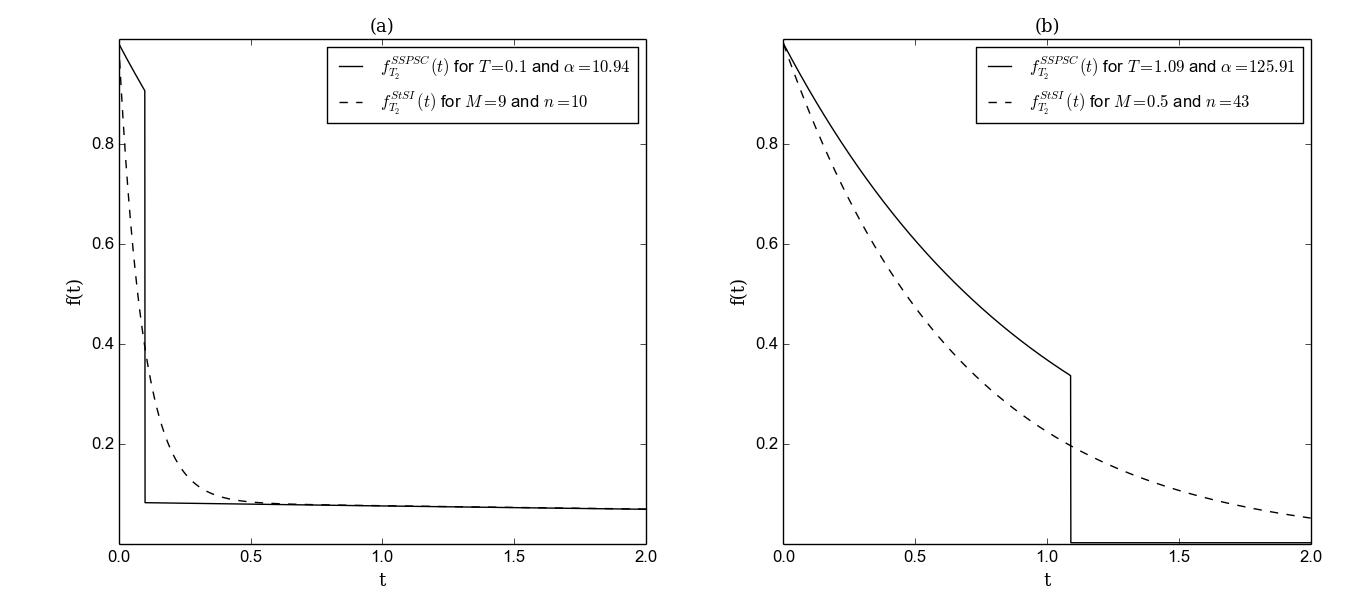}   
  \caption{Density of $T_2$ under the SSPSC and StSI models. Two sets of parameter values (panels (a) and (b), respectively) were chosen on the basis that expectations and variances were close. Panel (a): Density for the SSPSC model with $T=0.1$ and $\alpha=10.94$, and for the StSI model with $M=9$ and $n=10$. For this set of parameters we have $E\left(T_2^{SSPSC}\right)= 9.994$,  and $E\left(T_2^{StSI}\right)= 10$, $Var\left(T_2^{SSPSC}\right)=118.7$ and $Var\left(T_2^{StSI}\right)=118.0$. Panel (b): The same, but for $T=1.09$ and $\alpha=125.91$, and for $M=0.5$ and $n=43$. The corresponding expectations and variances are $E\left(T_2^{SSPSC}\right)= 42.997$,  and $E\left(T_2^{StSI}\right)= 43$, $Var\left(T_2^{SSPSC}\right)=8905$ and $Var\left(T_2^{StSI}\right)=8905$.}
  \label{fig:DensityCompare}
\end{figure}

The near-identical values obtained for the expectation and variance under the two models explains why it may be difficult to separate models of population size change from models of population structure when the number of independent genetic markers is limited. However, the differences between the distributions of coalescence times under the two models suggest that we can probably go further and identify one model from another. For instance, figure \ref{fig:DensityCompare} shows that even in cases where the first two moments are near-identical ($T=0.1$ and $\alpha=10$ versus $M=7$ and $n=9$), it should be theoretically possible to distinguish them. This is exactly what we aim to do in the next section. In practice, we will assume that we have a sample of $n_L$ independent $T_2$ values (corresponding to $n_L$ independent \textit{loci}) and will use these $T_2$ values to (i) estimate the parameter values that best explain this empirical distribution under the two models of interest, (ii) use a statistical test to compare the empirical distribution with the expected distribution for the ML estimates and reject (or not) one or both of the models. For simplicity, and to make it easier to read, we will often use the term \textit{loci} in the rest of the manuscript when we want to mention the number of independent $T_2$ values.

%////////////////Distribution_Moments/////////////////

%////////////////Parameter_Estimation_Model_Choice/////////////////
\subsection{Model choice and parameter estimation}

\newtheorem{thm}{Theorem}[section]
\newtheorem{lem}[thm]{Lemma}
\newtheorem{prop}[thm]{Proposition}
\newtheorem{corollary}[thm]{Corollary}

\newenvironment{proof}[1][Proof]{\begin{trivlist}
\item[\hskip \labelsep {\bfseries #1}]}{\end{trivlist}}
\newenvironment{definition}[1][Definition]{\begin{trivlist}
\item[\hskip \labelsep {\bfseries #1}]}{\end{trivlist}}
\newenvironment{example}[1][Example]{\begin{trivlist}
\item[\hskip \labelsep {\bfseries #1}]}{\end{trivlist}}
\newenvironment{remark}[1][Remark]{\begin{trivlist}
\item[\hskip \labelsep {\bfseries #1}]}{\end{trivlist}}

\subsubsection{General principle and parameter combinations:}
\label{subsec:Descrip_process}

Given a sample $(t_1,...,t_{n_L})$ of $n_L$ independent observations of the random variable $T_2$, we propose a parameter estimation procedure and a goodness-of-fit test to determine whether the observed distribution of the $T_2$ values is significantly different from that expected from the theoretical $T^{SSPSC}_2$ or $T^{StSI}_2$ distributions. This sample can be seen as a set of $T_2$ values obtained or estimated from $n_L$ independent loci. We took a Maximum Likelihood (ML) approach to estimate the parameters $(T,\alpha)$ and $(M,n)$ under the hypothesis that the $n_L$-sample was generated under the $T^{SSPSC}_2$ and the $T^{StSI}_2$ distributions, respectively (see Supplementary materials for the details of the estimation procedure). The ML estimates $(\widehat{T},\widehat{\alpha})$ and $(\widehat{M},\widehat{n})$ were then used to define $T^{SSPSC}_2$ or $T^{StSI}_2$ reference distributions. The Kolmogorov-Smirnov ($KS$) test which allows to compare a sample with a reference distribution was then used to determine whether the observed $n_L$ sample could have been generated by the respective demographic models. In other words this allowed us to reject (or not) the hypothesis that the  $(t_1,...,t_{n_L})$ sample was a realization of the reference distributions ($T_2^{StSI}$ or $T_2^{SSPSC}$). Note that the estimation procedure and the $KS$ test were performed on independent sets of $T_2$ values. We thus simulated twice as many $T_2$ values as needed ($2n_L$ instead of $n_L$).  With real data that would require that half of the loci be used to estimate  $(\widehat{T},\widehat{\alpha})$ and $(\widehat{M},\widehat{n})$, whereas the other half would be used to perform the $KS$ test.

We expect that if the estimation procedure is accurate and if the \textit{KS} test is performing well we should reject the SSPSC (respectively, the StSI) model when the data were simulated under the StSI (resp., the SSPSC) model. On the contrary we should not reject data simulated under the SSPSC (resp., the StSI) model when they were indeed simulated under that model. To validate our approach we used $(t_1,...,t_{2n_L})$ data sampled from the two $T_2$ distributions and quantified how the estimation procedure and the $KS$ test performed. In order to do that, we varied the parameter values ($(T, \alpha)$ and $(M, n)$) for various $2n_L$ values as follows. For $T$ and $\alpha$ we used all 36 pairwise combinations between these two sets of values (0.1, 0.2, 0.5, 1, 2, 5), and (2, 4, 10, 20, 50, 100), respectively. For $M$ and $n$ we used all the 48 combinations between the following values (0.1, 0.2, 0.5, 1, 5, 10, 20, 50) and (2, 4, 10, 20, 50, 100), respectively. For $2n_L$ we used the following values (40, 100, 200, 400, 1000, 2000, 20000). Altogether we tested $588$ combinations of parameters and number of loci. For each $2n_L$ value and for each parameter combination $(T, \alpha)$ (or $(M, n)$) we realized $100$ independent repetitions of the following process. We first simulated a sample of $2n_L$ values using the $pdfs$ of the SSPSC (resp. StSI) model with $(T, \alpha)$ (resp. $(M, n)$). We then used the first $n_L$ values to obtain the ML estimates $(\widehat{T}, \widehat{\alpha})$ for the SSPSC model and $(\widehat{M}, \widehat{n})$ for the StSI model. Then, we performed a $KS$ test using a $0.05$ threshold on the second half of the simulated data (\textit{i.e.} $n_L$ values) with each of the theoretical distributions defined by the estimated parameters. Finally, after having repeated this process $100$ times we recorded all estimated parameters and counted the number of times we rejected the SSPSC and StSI models for each parameter combination and each $2n_L$ value.

\subsubsection{Maximum likelihood estimation in the SSPSC case:}

We know from section \ref{eq:pdf_T2_SSPCS} the $pdf$ of the coalescence time in the SSPSC model of two genes. We can thus write the likelihood function for any couple of parameters $(\alpha, T)$, given one observation $t_i$ as:

\begin{equation*}
  \label{eq:LlkBot_ti}
    \mathbb{L}_{t_i}(\alpha, T) = \frac{1}{\alpha}e^{-T-\frac{1}{\alpha}(t_i-T)}\mathbb{I}_{[0, t_i[}(T) + e^{-t_i}\mathbb{I}_{]t_i,+\infty[}(T).
\end{equation*}

Given $n_L$ independent values $t = (t_1, t_2, ..., t_{n_L})$, the likelihood is:

\begin{equation*}
  \label{eq:LlkBot}
  \mathbb{L}_{SSPSC}(\alpha, T) = \displaystyle\prod_{i=1}^{n_L}\mathbb{L}_{t_i}(T, \alpha),
\end{equation*}

\noindent and taking the \textit{log} it gives:

\begin{equation*}
  \label{eq:LogLkBot}
  \log(\mathbb{L}_{SSPSC}(\alpha, T)) = \displaystyle\sum_{i=1}^{n_L}\log(\mathbb{L}_{t_i}(\alpha, T)).
\end{equation*}

\begin{lem}
\label{Lem:CriticalPoints_Llk}

Given a set of $n_L$ independent observations $\{t_1, t_2, ..., t_{n_L}\}$, the log-likelihood function $\log(\mathbb{L}_{SSPSC})$ has no critical points in $\mathbb{R}^2$.
\end{lem}

For the proof and some comments, see Supplementary Materials.

As a consequence of this lemma, we take $\displaystyle (\hat{\alpha}, \hat{T}) = argmax_{a\in\{1, ..., n_L\}}\{\log(\mathbb{L}_{SSPSC}(m_a))\}$ as the Maximum Likelihood estimates, where

\begin{equation*}
  \label{eq:CritPoints}
  \begin{split}
    m_a=\left(\frac{\displaystyle\sum_{i=1}^{n_L}t_i\mathbb{I}_{t_a<t_i} - Kt_a}{K+1}, t_a\right) &, a \in \{1, 2, ..., n_L\}.
  \end{split}
\end{equation*}

with

\begin{equation*}
 K=\displaystyle\sum_{i=1}^{n_L}\mathbb{I}_{t_i < t_a}
\end{equation*}

\subsubsection{Maximum likelihood estimation in the StSI case:}

Under the StSI model the expression of the critical points is not analytically derived. We know from section \ref{eq:pdf_T2_StSI} the $pdf$ of coalescence times for two genes. Given $n_L$ independent values $t=(t_1, t_2, ..., t_{n_L})$ we can compute the log-likelihood function for any set of parameters $(n,M)$ as:

\begin{equation*}
  \label{eq:log-likelihood_StSI}
  \log(\mathbb{L}_{StSI}(n, M)) = \displaystyle\sum_{i=1}^{n_L}\log(ae^{-\alpha t}+(1-a)e^{-\beta t}))
\end{equation*}

We used the Nelder-Mead method \cite{nelder1965simplex} implementation of \textit{scipy} \cite{Scipy2001} to find numerically an approximation to the maximum of the likelihood function. This method returns a pair of real numbers $(\hat{n}, \hat{M})$. Since $n$ should be an integer we kept either $\lfloor \hat{n} \rfloor$ or $\lfloor \hat{n} \rfloor +1$, depending on which had the largest log-likelihood value.

\newpage
\section{Results}

\begin{figure}
  \centering
  \includegraphics[width=\textwidth]{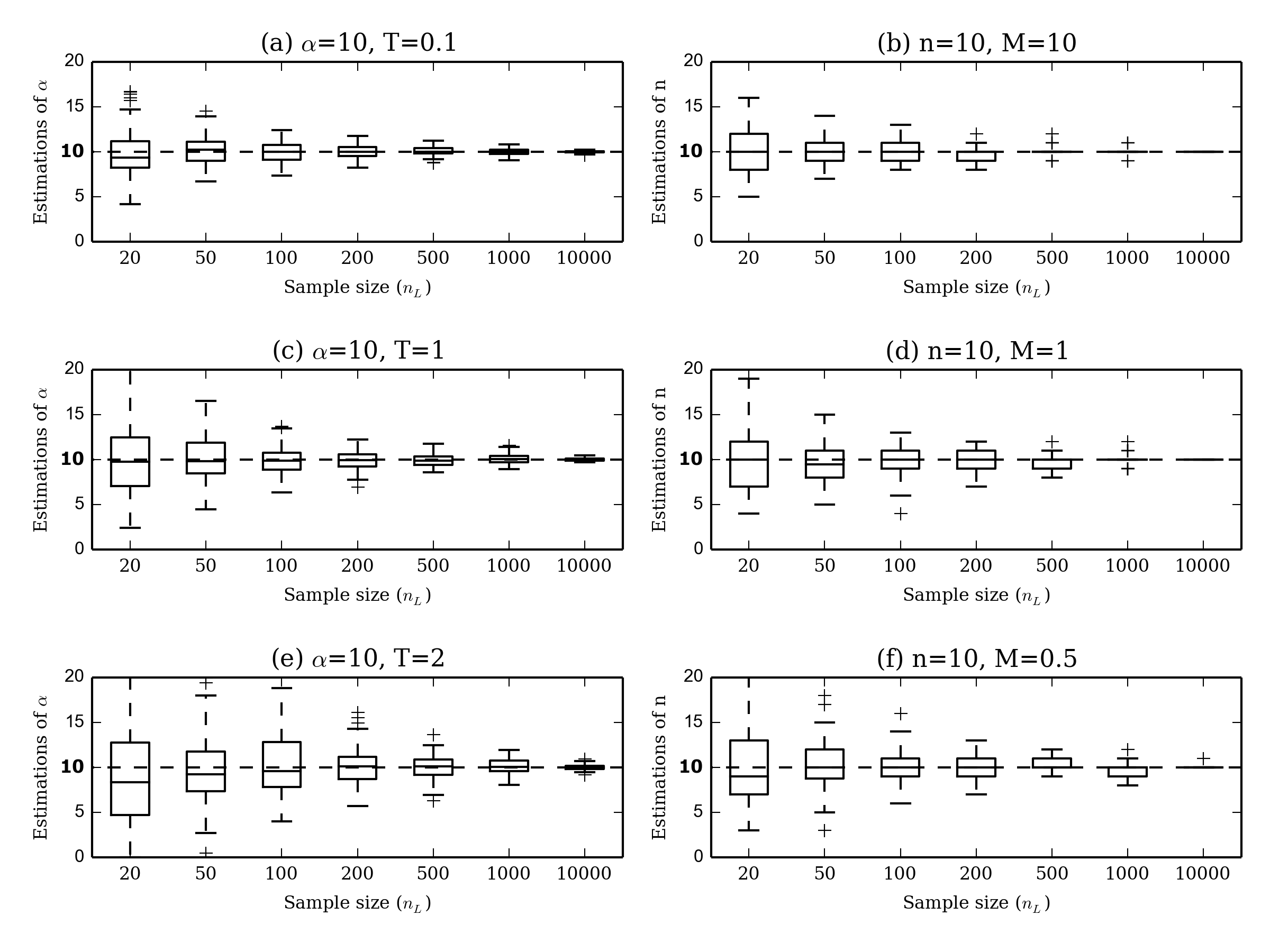}
  \caption{Estimation of $\alpha$ and $n$. Panels (a), (c) and (e): Estimation of $\alpha$ under the SSPSC model for different sample sizes and $T$ values. Simulations performed with $\alpha=10$ and $T=(0.1, 1, 2)$. Panels (b), (d) and (f): Estimation of $n$ under the StSI model for different sample sizes and $M$ values. Simulations performed with $n=10$ and $M=(10, 1, 0.5)$.}
  \label{fig:Estim_Acc_alpha_n_10}
\end{figure}

Figure \ref{fig:Estim_Acc_alpha_n_10} shows, for various values of $n_L$, the results of the estimation of $\alpha$ (panels (a), (c), and (e), for simulations assuming $\alpha=10$ and $T=(0.1, 1, 2)$, respectively ; see Supplementary Material for the other values) and the estimation of $n$ (panels (b), (d), and (f) for simulations with $n=10$ and $M=(10, 1, 0.5)$, respectively; see Supplementary Material for the other values, corresponding to 26 figures and 168 panels). The first thing to notice is that both $\alpha$ and $n$ are increasingly well estimated as $n_L$ increases. This is what we expect since $n_L$ represents the amount of information (the number of $T_2$ values or independent loci.) The second thing to note is that the two parameters are very well estimated when we use $10,000$ values of $T_2$. This is particularly obvious for $n$ compared to $\alpha$, probably because $n$ must be an integer, whereas $\alpha$ is allowed to vary continuously. For instance, for most simulations we find the exact $n$ value (without error) as soon as we have more than $1000$ loci. However, we should be careful in drawing very general rules. Indeed, when fewer $T_2$ values are available (\textit{i.e.} fewer independent loci), the estimation precision of both parameters depends also on $T$ and $M$, respectively. Interestingly, the estimation of $\alpha$ and $n$ are remarkable even when these parameters are small. This means that even ``mild'' bottlenecks may be very well quantified (see for instance the Supplementary materials for $\alpha=2$, $T$ values between $0.1$ and $1$ when we use only 1000 loci). We should also note that when the bottleneck is very old ($T=5$) the estimation of the parameters is rather poor and only starts to be reasonable and unbiased for $n_L=10,000$. This is not surprising since the expected $T_{MRCA}$ is $1$. Under the SSPSC model most genes will have coalesced by $t=5$, and should therefore exhibit $T_2$ values sampled from a stationary population (\textit{i.e.} $\alpha=1$). As the number of loci increases, a small proportion will not have coalesced yet and will then provide information on $\alpha$. The expected proportion of genes that have  coalesced by $T=5$ is 0.993.  

\begin{figure}
  \centering
  \includegraphics[width=\textwidth]{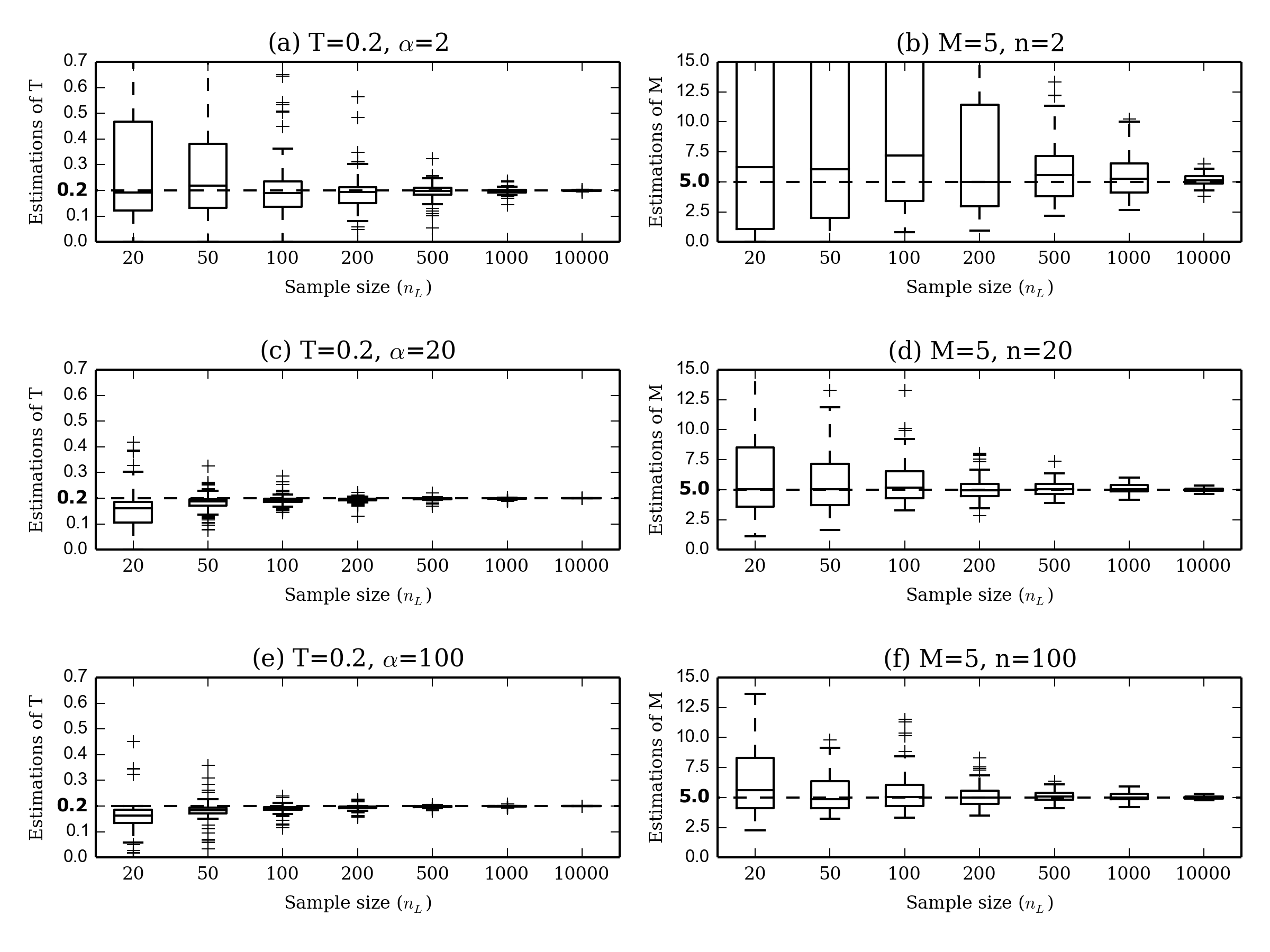}
  \caption{Estimation of $T$ and $M$. Panels (a), (c), (e)): Estimation of $T$ under the SSPSC model for different sample sizes and values of $\alpha$. Simulations performed with $\alpha=(2, 20, 100)$ and $T=0.2$. Panels (b), (d), (f): Estimation of $M$ under the StSI model for different sample sizes and values of $n$. Simulations performed with $n=(2, 20, 100)$ and $M=5$.}
  \label{fig:Estim_Acc_T_M}
\end{figure}

Figure \ref{fig:Estim_Acc_T_M} shows for various values of $n_L$ the results of the estimation of $T$ (panels (a), (c), and (e), for simulations assuming $T=0.2$ and $\alpha=(2, 20, 100)$, respectively; see Supplementary Material for the other values) and the estimation of $M$ (panels (b), (d), and (f), for simulations with $M=20$ and $n=(2, 20, 100)$, respectively; see Supplementary Material for the other values). As expected again, the estimates are getting better as $n_L$ increases. For the values shown here we can see that $T$, the age of the bottleneck is very well estimated even when $\alpha=2$ (for $n_L=10,000$). In other words, even a limited bottleneck can be very precisely dated. For stronger bottlenecks fewer loci (between 500 and 1000) are needed to still reach a high precision. This is particularly striking given that studies suggest that it is hard to identify bottlenecks with low $\alpha$ values \cite{Girod01052011}. Interestingly, the panels (b), (d) and (f) seem to suggest that it may be more difficult to estimate $M$ than $T$. As we noted above this observation should be taken with care. Indeed, $T$ and $M$ are not equivalent in the same way as $\alpha$ and $n$. This is why we chose to represent a value of $M$ such that $M=1/T$, and why one should be cautious in drawing general conclusions here. Altogether this and the previous figure show that it is possible to estimate with a high precision the parameters of the two models by using only $500$ or $1000$ loci from a single diploid individual. There are also parameter combinations for which much fewer loci could be sufficient (between $50$ and $100$).

In Figure \ref{fig:Bot_rejections} we show some results of the $KS$ test for the two cases (See the Supplementary Materials for the other parameter combinations). In the left-hand panels ((a), (c), and (e)) the data were simulated under the SSPSC model and we used the StSI model as a reference (\textit{i.e.} we ask whether we can reject the hypothesis that genetic data were generated under a structured model when they were actually generated under a model of population size change). In the right-hand panels ((b), (d) and (f)) the same data were compared using the SSPSC model as reference and we computed how often we rejected them using a $5\%$ rejection threshold. The left-hand panels exhibit several important features. The first is that, with the exception of $T_2=5$ we were able to reject the wrong hypothesis in $100\%$ of the cases when we used $10,000$ independent $T_2$ values. 

This shows that our estimation procedure (as we saw above in figures \ref{fig:Estim_Acc_alpha_n_10} and  \ref{fig:Estim_Acc_T_M}) and the $KS$ test are very powerful. The second feature is that for $T=5$, the test performs badly whatever the number of independent loci (at least up to $10,000$). This is expected since the expected $T_{MRCA}$ of two genes is $t=1$, and $99.3\%$ of the loci will have coalesced by $t=5$. This means that out of the $10,000$, only \textit{c.a.} 70 loci are actually informative regarding the pre-bottleneck population size. Another important feature of the left-hand panels is that the best results are generally obtained for $T= 1, 0.5$ and $2$, whichever the value of $\alpha$. This is in agreement with \citeN{Girod01052011} in that very recent population size changes are difficult to detect and quantify. The observation is valid for ancient population size changes as well. The right-hand panels are nearly identical, whichever $\alpha$ value we used  (see also Supplementary Materials), and whichever number of $T_2$ values we use. They all show that the $KS$ test always rejects a rather constant proportion of data sets. This proportion varies between $3$ and $15\%$, with a global average of $8.9\%$. Altogether our $KS$ test seems to be conservative. This is expected because for low $n_L$ values the estimation of the parameters will tend to be poor. Since the $KS$ test uses a reference distribution based on the estimated rather than the true values, it will reject the simulated data more often than the expected value of $5\%$.

Figure \ref{fig:Struct_rejections} is similar to Figure \ref{fig:Bot_rejections} but the data were simulated under the StSI model and the $KS$ test was performed first using the SSPSC model as a reference ((a), (c), (e)) and then using the StSI model as a reference ((b), (d), (f)). The left-hand panels ((a), (c), and (e)) show results when we ask whether we can reject the hypothesis that genetic data were generated under a population size change model when they were actually generated under a model of population structure. In the right-hand panels ((b), (d), and (f)) we computed how often we rejected the hypothesis that genetic data were generated under the StSI model when they were indeed generated under that model of population structure. Altogether, the left-hand panels suggest that the results are generally best when $M = (0.1, 0.2, 1)$, but that we get very good results for most values of $M$ when we have $10,000$ loci and can reject the SSPSC when they were actually generated under the StSI model. The right-hand panels show, as in Figure \ref{fig:Bot_rejections}, that for all the values of $n_L$ and $n$ we reject a rather constant proportion of data sets (between $5$ and $10\%$). Altogether the two previous figures (figures \ref{fig:Bot_rejections} and \ref{fig:Struct_rejections}) show that it is possible to identify the model under which genetic data were generated by using genetic data from a single diploid individual. 

Figure \ref{fig:Rel_Btw_Params} is divided in four panels showing the relationships between $T$ and $M$ (panels (b) and (d), for various values of $\alpha$ and $n$) and between $\alpha$ and $n$ (panels (a) and (c), for various values of $T$ and $M$). In each of the panels we simulated data under a model for specific parameter values represented on the x-axis, and estimated parameters from the other model, and represented the estimated value on the y-axis. Since we were interested in the relationship between parameters (not in the quality of the estimation, see above), we used the largest $n_L$ value and plotted the average of 100 independent estimation procedures. In panel (a) we simulated a population size change (SSPSC) for various $T$ values (represented each by a different symbol) and several values of $\alpha$ on the x-axis. We then plotted the estimated value of $\widehat{n}$ for each case (\textit{i.e.} when we assume that the data were generated under the StSI model). We find a striking linear relationship between these two parameters conditional on a fixed $T$ value. For instance, a population bottleneck by a factor $50$ that happened $N_0$ generations ago ($T=1$) is equivalent to a structured population with $\widehat{n}\approx22$ islands (and $\widehat{M}\approx 0.71$). Panel (c) is similar and shows how data simulated under a structured population generates specific parameters of population bottlenecks. Panels (b) and (d) show the relationship between $T$ and $M$. We have plotted as a reference the curve corresponding to $y=1/x$. As noted above and shown on this graph, this relationship is only approximate and depends on the value of $\alpha$ and $n$. Altogether, this figure exhibits the profound relationships between the model parameters. They show that the qualitative relationships between $\alpha$ and $n$, and between $T$ and $1/M$ discussed above are indeed real but only correct up to a correcting factor. Still this allows us to identify profound relationships between population structure and population size change.

\begin{figure}
  \centering
  \includegraphics[width=\textwidth]{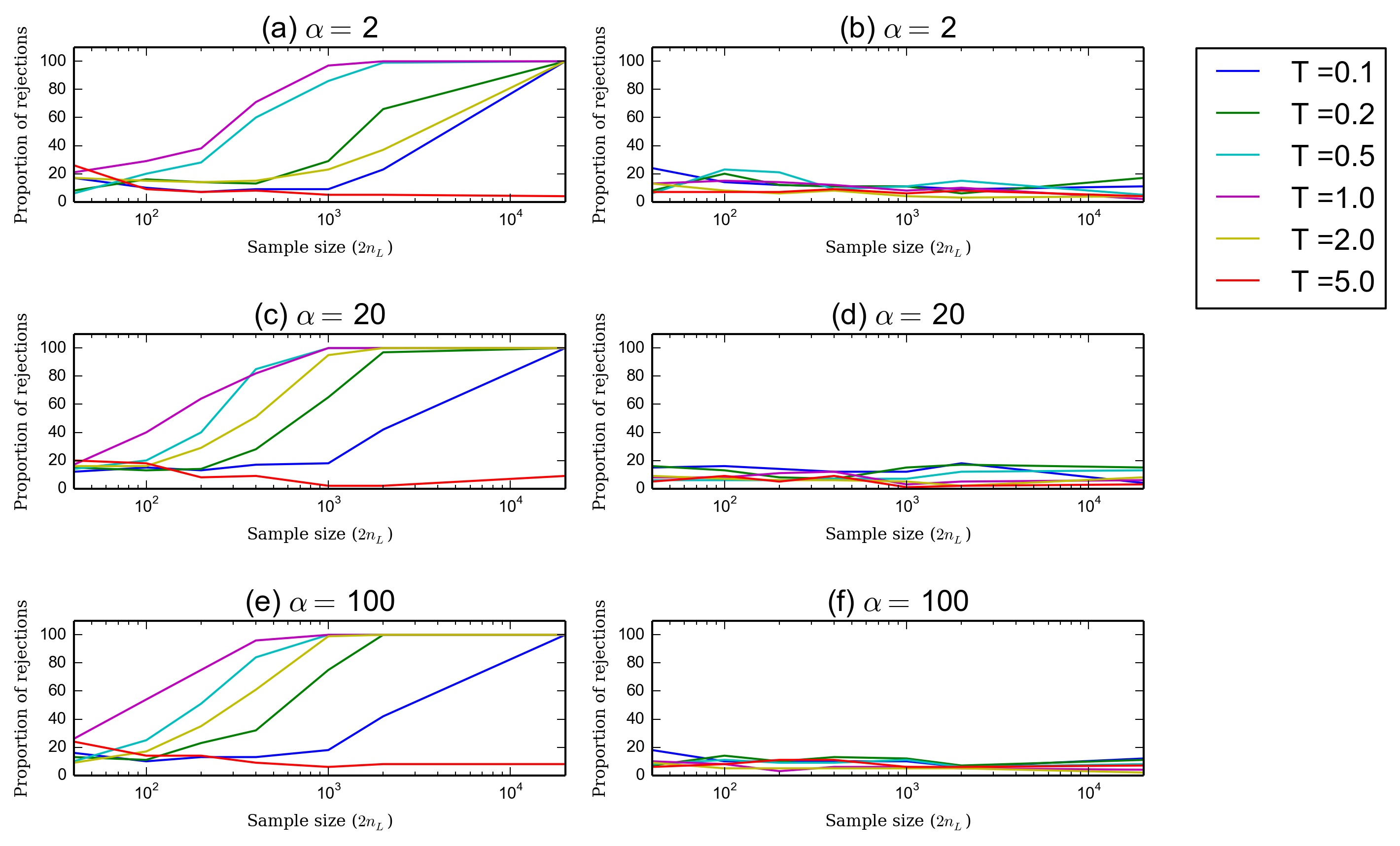} 
  \caption{Proportion of rejected data sets simulated under the SSPSC model. Panels (a), (c) and (e): the reference model is the StSI model. Panels (b), (d), and (f): the reference model is the SSPSC, \textit{i.e.} the model under which the data were simulated. Note that for the abscissa we used $2n_L$ instead of $n_L$ because in order to perform the $KS$ test it is necessary to first estimate the parameters using $n_L$ loci and then an independent set of $n_L$ values of $T_2$.}
  \label{fig:Bot_rejections}
\end{figure}

\begin{figure}
  \centering
  \includegraphics[width=\textwidth]{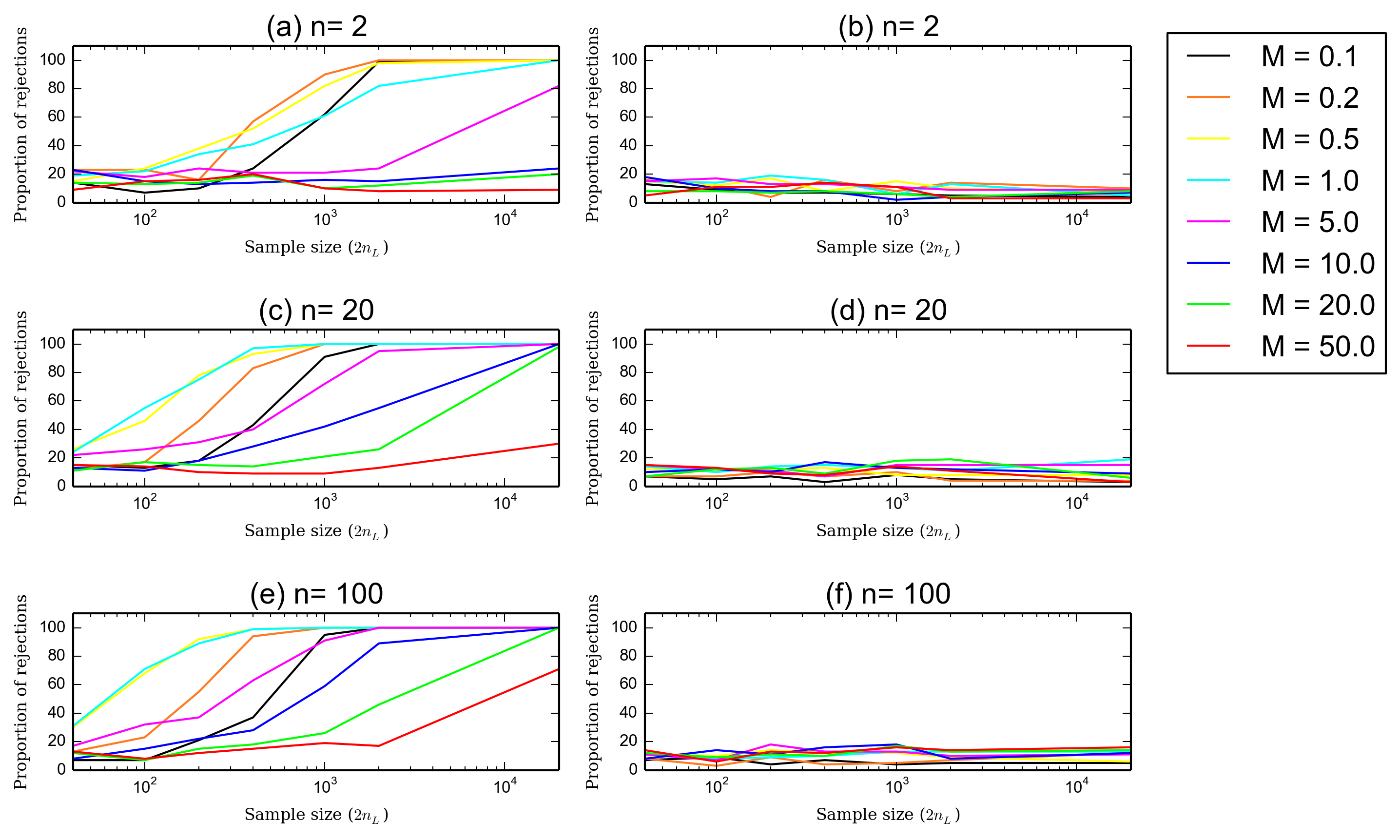} 
  \caption{Proportion of rejected data sets simulated under the StSI model. Panels (a), (c), and (e): the reference model is the SSPSC.  Panels (b), (d), and (f): the reference model is the StSI model, \textit{i.e.} the model under which the data were simulated. Note that for the abscissa we used $2n_L$ instead of $n_L$ because in order to perform the $KS$ test it is necessary to first estimate the parameters using $n_L$ loci and then an independent set of $n_L$ values of $T_2$.}
  \label{fig:Struct_rejections}
\end{figure}

\begin{figure}
  \centering
  \includegraphics[width=\textwidth]{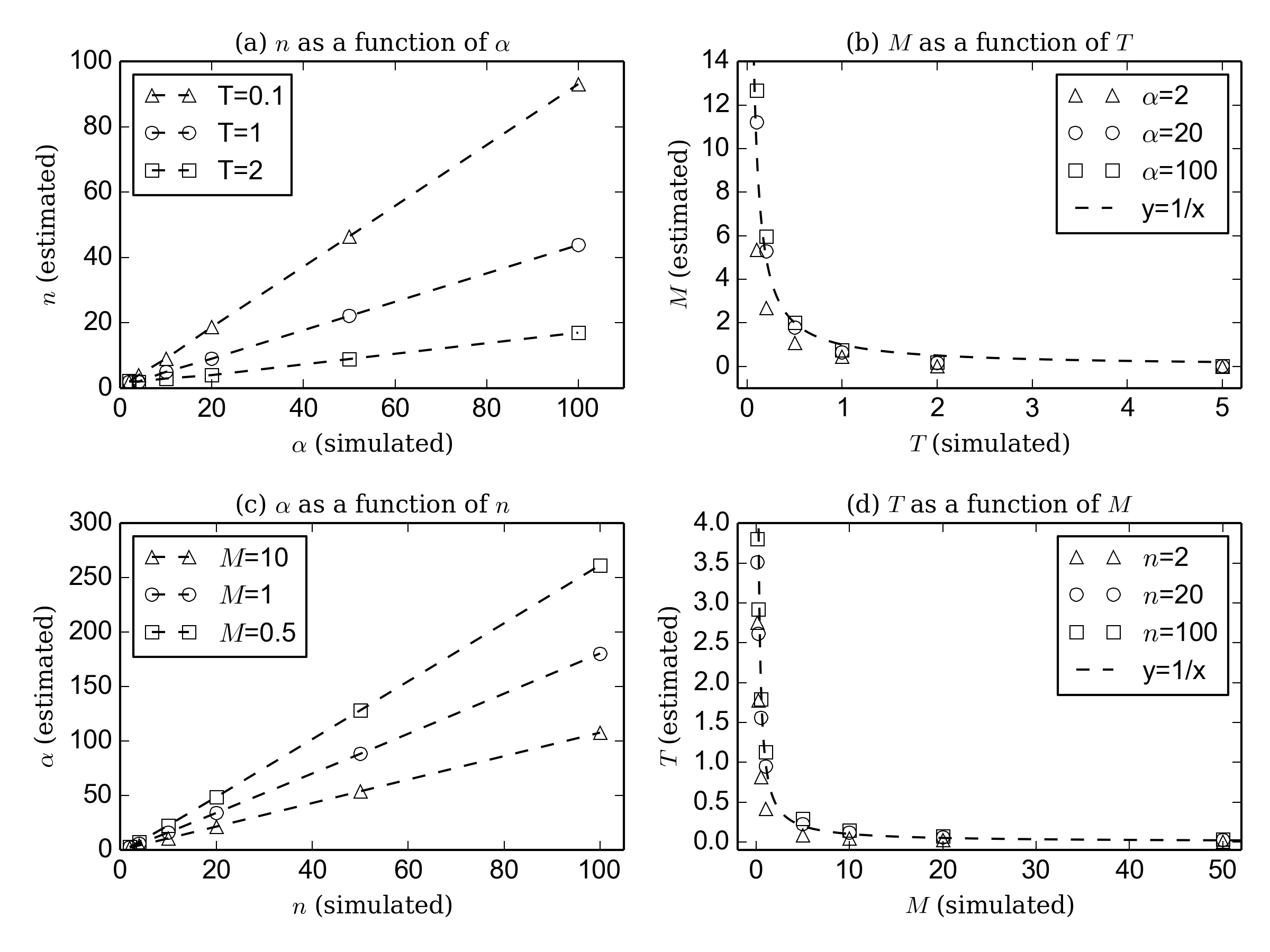}
  \caption{Relationships between parameters of the models}
  \label{fig:Rel_Btw_Params}
\end{figure}

%////////////////Parameter_Estimation_Model_Choice/////////////////

%////////////////Discussion/////////////////
\section{Discussion}

In this study we have analysed the distribution of coalescence times under two simple demographic models. We have shown that even though these demographic models are strikingly different (Figure 1) there is always a way to find parameter values for which both models will have the same first two moments (Figure \ref{fig:ExpVarCompare}). We have also shown that there are intrinsic relationships between the parameters of the two models (Figure \ref{fig:Rel_Btw_Params}). However, and this is a crucial point, we showed that the distributions were different and could therefore be distinguished. Using these distributions we developed a $ML$ estimation procedure for the parameters of both models $(\widehat{T},\widehat{\alpha})$ and $(\widehat{M},\widehat{n})$ and showed that the estimates are accurate, given enough genetic markers. Finally, we showed that by applying a simple $KS$ test we were able to identify the model under which specific data sets were generated. In other words, we were able to determine whether a bottleneck signal detected in a particular data set could actually be caused by population structure using genetic data from a single individual. The fact that a single individual provides enough information to estimate demographic parameters is in itself striking (see in particular the landmark paper by \citeN{li2011inference}), but the fact that one individual (or rather sometimes as few as $500$ or $1000$ loci from that one individual) potentially provides us with the ability to identify the best of two (or more) models is even more remarkable. The PSMC (pairwise sequentially markovian coalescent) method developed by \citeN{li2011inference} reconstructs a theoretical demographic history characterized by population size changes, assuming a single non structured population. Our study goes further and shows that it could be possible to test whether the signal identified by the PSMC is due to actual population size changes or to population structure. However, models putting together these two scenarios have being proposed. In \citeN{wakeley1999nonequilibrium}, a model considering a structured population who went through a bottleneck in the past is developed, allowing to estimate the \textit{time} and the \textit{ratio} of the bottleneck. Moreover, \citeN{wakeley1999nonequilibrium} discussed the idea that, in structured populations, changes in the migration rate and/or the number of islands (demes) can shape the observed data in the same way that effective population size changes do. Hence, we think that our work could be helpful to the aim of setting these two scenarios apart in order to detect (for example) false bottleneck signals. Nevertheless, while our study provides several new results, there are still several important issues that need to be discussed and much progress that can still be made.

\subsection{$T_2$ and molecular data}
\paragraph{}
The first thing to note is that we assume, throughout our study, that we have access to the coalescence times $T_2$. In real data sets, this is never the case and the $T_2$ are rarely estimated from molecular data. While this may seem as a limitation, we should note that the recent method of \citeN{li2011inference}, that uses the genome sequence of a single individual to infer the demographic history of the population it was sampled from, actually estimates the distribution of $T_2$ values. In its current implementation the PSMC software does not output this distribution but it could be modified to do it. Note however, that the PSMC should only be able to provides a discretized distribution in the form of a histogram with classes defined by the number of time periods for which population size estimates are computed. In any case, this suggests that it is in theory possible to use the theoretical work of Li and Durbin to generate $T_2$ distributions, which could then be used with our general approach.
Moreover, it should be possible to use the theory developed here to compute, conditional of the $T_2$ distribution, the distribution of several measures of molecular polymorphism. For instance, under an infinite site mutation model it is in principle possible to compute the distribution of the number of differences between pairs of non recombining sequences for the two demographic models analysed here. Similarly, assuming a single stepwise mutation model it should be possible to compute the distribution of the number of repeat differences between two alleles conditional on $T_2$. However, we must add here that while it is easy to use these distributions to simulate genetic data (rather than $T_2$ values) it is not straightforward to then use the genetic data to estimate the models' parameters and apply a test to identify the model under which the data were simulated. This would probably require a Khi-2 test since discrete rather continuous distributions would be compared. This is an issue that would deserve a full and independent study.

\subsection{Demographic models}
\paragraph{}
In our study we limited ourselves to two simple but widely used models. It would thus be important to determine the extent to which our approach could apply to other demographic models. The n-island or StSI model is a widely used model and it was justified to use it here. One of its strongest assumptions is that migration is identical between all demes. This is likely to be problematic for species with limited vagility. In fact, for many species a model where migration occurs between neighbouring populations such as the stepping-stone is going to be more likely. At this stage it is unclear whether one could derive analytically the $pdf$ of $T_2$ for a stepping-stone model. The work by \citeN{herbots1994stochastic} suggests that it may be possible to compute it numerically by inversing the Laplace transform derived by this author. This work has not been done to our knowledge and would still need to be done. Interestingly, this author has also shown that it is in principle possible to derive analytically the $pdf$ of $T_2$ in the case of a two-island model with populations of different sizes. Again, this still needs to be done.

The SSPSC model has also been widely used \cite{rogers1992population} and represents a first step towards using more complex models of population size change. For instance, the widely used method of \citeN{beaumont1999detecting} to detect, date and quantify population size changes \cite{goossens2006genetic,quemere2012genetic,salmona2012signature} assumes either an exponential or a linear population size change. It should be straightforward to compute the $pdf$ of $T_2$ under these two models because, as we explained above, the coalescent theory for populations with variable size has been very well studied \cite{donnelly1995coalescents,tavare2004part} and it is possible to write the $pdf$ of $T_2$ for any demographic history involving any type of population size changes. To go even further one could in principle use the history reconstructed by the PSMC \cite{li2011inference} as the reference model of population size change and compare that particular demographic history to an n-island model using our results, and our general approach. 
At the same time, we should note that for complex models of population size change, including relatively simple ones such as the exponential model of \citeN{beaumont1999detecting}, it is not straightforward to compute the number of differences between non recombining sequences. Significant work would probably be needed to apply the general approach outlined in our study to specific demographic models. But we believe that the possibilities opened by our study are rather wide and should provide our community with new interesting problems to solve in the next few years.

\subsection{Comparison with previous work and generality our of results}
\paragraph{}
The present work is part of a set of studies aimed at understanding how population structure can be mistaken for population size change and at determining whether studies identifying population size change are mistaken or valid \cite{chikhi2010confounding,heller2013confounding,paz2013demographic}. It is also part of a wider set of studies that have recognised in the last decade the importance of population structure as potential factor biasing inference of demographic \cite{leblois2006genetics,stadler2009impact,peter2010distinguishing,chikhi2010confounding,heller2013confounding,paz2013demographic} or selective processes \cite{currat2006comment}. Here we demonstrated that it is indeed possible to separate the SSPSC and StSI models. While we believe that it is an important result, we also want to stress that we should be cautious before extending these results to any set of models, particularly given that we only use the information from $T_2$. Much work is still needed to devise new tests and estimation procedures for a wider set of demographic models and using more genomic information, including recombination patterns as in the PSMC method \cite{li2011inference}. Beyond the general approach outlined here we would like to mention the study of \citeN{peter2010distinguishing} who also managed to separate one structure and one PSC (\textit{Population Size Change}) model. These authors used an ABC approach to separate a model of  exponential PSC from a model of population structure similar to the StSI model. Their structured model differs from ours by the fact that it is not an equilibrium model. They assumed that the population was behaving like an n-island model in the recent past, until T generations in the past, but that before that time, the ancestral population from which all the demes derived was not structured. When T is very large their model is identical to the StSI, but otherwise it may be quite different. The fact that they managed to separate the two models using an ABC approach is promising as it suggests that there is indeed information in the genetic data for models beyond those that we studied here. We can therefore expect that our approach may be applied to a wider set of models. We should also add that in their study these authors use a much larger sample size (25 diploid individuals corresponding to 50 genes). They used a maximum of 200 microsatellites which corresponds therefore to 10,000 genotypes, a number very close to the maximum number we used here. Altogether our study provides new results and opens up new avenues of research for the distribution of coalescent times under complex models.

\subsection{Sampling and population expansions}
\paragraph{}
Recent years have also seen an increasing recognition of the fact that the sampling scheme together with population structure may significantly influence demographic inference \cite{wakeley1999nonequilibrium,stadler2009impact,chikhi2010confounding,heller2013confounding}. For instance, \citeN{wakeley1999nonequilibrium} showed that in the n-island model genes sampled in different demes will exhibit a genealogical tree similar to that expected under a stationary Wright-Fisher model. Since our work was focused on $T_2$ we mostly presented our results under the assumption that the two genes of interest were sampled in the same deme. For diploids this is of course a most reasonable assumption. However, we should note that the results presented above also allow us to express the distribution of $T_2$ when the genes are sampled in different demes. We did not explore this issue further here, but it would be important to study the results under such conditions. Interestingly we find that if we assume that the two genes are sampled in two distinct demes, we can detect population expansions rather than bottlenecks. This could happen if we considered a diploid individual whose parents came from different demes. In that case, considering the two genes sampled in the deme where the individual was sampled would be similar to sampling his two parental genes in two different demes. Interestingly, this has also been described by \citeN{peter2010distinguishing} who found that when the 25 individuals were sampled in different demes, they would detect population size expansions rather than bottlenecks. Our results are therefore in agreement with theirs. Similarly \citeN{heller2013confounding} also found that some signals of population expansion could be detected under scattered sampling schemes.

\subsection{Conclusion: islands within individuals}
\paragraph{}
To conclude, our results provide a general framework that can in theory be applied to whole families of models. We showed for the first time that genomic data from a single individual could be used to estimate parameters that have to our knowledge never been estimated. In particular we showed that we were able to estimate the number of islands (and the number of migrants) in the StSI model. This means that one can in principle use genomic data from non model organisms to determine how many islands make up the metapopulation from which one single individual was sampled. This is of course true as well for model organisms but it is particularly meaningful for species for which the number of individuals with genomic data is limited. Our ability to estimate $n$ is one of the most powerful results of our study. While such estimates should not be taken at face value, they surely should be obtained across species for comparative analyses. Also, during the last decade there has been a major effort to use programs such as STRUCTURE \cite{pritchard2000inference} to estimate the number of "subpopulations" within a particular sample. Our work suggests that we might in principle provide additional results with only one individual. It is important to stress though that the answer provided here is very different from those obtained with STRUCTURE and similar methods and programs \cite{pritchard2000inference,guillot2005geneland,chen2007bayesian,corander2004baps}. We do not aim at identifying the populations from which a set of individuals come. Rather we show that his/her genome informs us on the whole set of populations. In other words, even though we assume that there are $n$ populations linked by gene flow, we show that each individual, is somehow a genomic patchwork from this (poorly) sampled metapopulation. We find these results reassuring, in an era where genomic data are used to confine individuals to one population and where division rather than connectivity is stressed.

%////////////////Discussion/////////////////

%////////////////Acknowledgements/////////////////
\section*{Acknowledgements}

We are grateful to S. Boitard and S. Grusea for numerous and fruitful discussions and input throughout the development of this work. We thank I. Paz, I. Pais, J. Salmona, J. Chave for discussion and helpful comments that improved the clarity of the text. We thank J. Howard and the Fundação Calouste Gulbenkian for their continuous support. We are grateful to the genotoul bioinformatics platform Toulouse Midi-Pyrenees for providing computing and storage resources. This work was partly performed using HPC resources from CALMIP (Grant 2012 - projects 43 and 44) from Toulouse, France. This study was funded by the Fundação para a Ciência e Tecnologia (ref. PTDC/BIA- BIC/4476/2012), the Projets Exploratoires Pluridisciplinaires (PEPS 2012 Bio-Maths-Info) project, the LABEX entitled TULIP (ANR-10-LABX-41) as well as the Pôle de Recherche et d'Enseignement Supérieur (PRES) and the Région Midi-Pyrénées, France.

%////////////////Acknowledgements/////////////////

\bibliography{Bibliography}
\bibliographystyle{mychicago}

\end{document}

\maketitle

\newtheorem{thm}{Theorem}[section]
\newtheorem{lem}[thm]{Lemma}
\newtheorem{prop}[thm]{Proposition}
\newtheorem{corollary}[thm]{Corollary}

\newenvironment{proof}[1][Proof]{\begin{trivlist}
\item[\hskip \labelsep {\bfseries #1}]}{\end{trivlist}}
\newenvironment{definition}[1][Definition]{\begin{trivlist}
\item[\hskip \labelsep {\bfseries #1}]}{\end{trivlist}}
\newenvironment{example}[1][Example]{\begin{trivlist}
\item[\hskip \labelsep {\bfseries #1}]}{\end{trivlist}}
\newenvironment{remark}[1][Remark]{\begin{trivlist}
\item[\hskip \labelsep {\bfseries #1}]}{\end{trivlist}}

\section{Proof of the lemma}
%\subsection{Proof of the Lemma }
\label{Proof:LemCriticalPoints_Llk}

\begin{lem}
Given set of $n_L$ independent observations $(t_1, t_2, ..., t_{n_L})$, the log-likelihood function $\log(\mathbb{L}_{SSPSC})$ has no critical points in $\mathbb{R}^2$.

\end{lem}

\begin{proof}

Given $n_L$ independent values $t = (t_1, t_2, ..., t_{n_L})$, the likelihood is:

\begin{equation*}
  \label{eq:LlkBot}
  \mathbb{L}_{SSPSC}(\alpha, T) = \displaystyle\prod_{i=1}^{n_L}\mathbb{L}_{t_i}(T, \alpha),
\end{equation*}

and taking the log:

\begin{equation}
\label{eq:log-Likelihood_function}
  \log(\mathbb{L}_{SSPSC}(\alpha, T)) = \displaystyle\sum_{i=1}^{n_L}\log(\mathbb{L}_{t_i}(\alpha, T)),
\end{equation}

\noindent where

\begin{equation}
\label{eq:Likelihood_function}
 \mathbb{L}_{t_i}(T, \alpha) = \frac{1}{\alpha}e^{-T-\frac{1}{\alpha}(t_i-T)}\mathbb{I}_{T\leq t_i} + e^{-t_i}\mathbb{I}_{T>t_i}
\end{equation}

First note that:
\begin{itemize}
\item For $T>t_i$, $\mathbb{L}_{t_i}(\alpha, T) = e^{-t_i}$ is \textbf{constant} (with respect to $(\alpha,T)$).
\item If $\alpha\neq 1$ ($\alpha=1$ means there is no change in the population's size) then
$\mathbb{L}_{t_i}(T, \alpha)$ has a discontinuity at $T=t_i$.
\end{itemize}

As we are interested in the case $\alpha\neq 1$, the log-likelihood function has discontinuities at each $t_i$, $i=1\ldots n_L$.

\begin{figure}
  \centering
  \includegraphics[width=\textwidth]{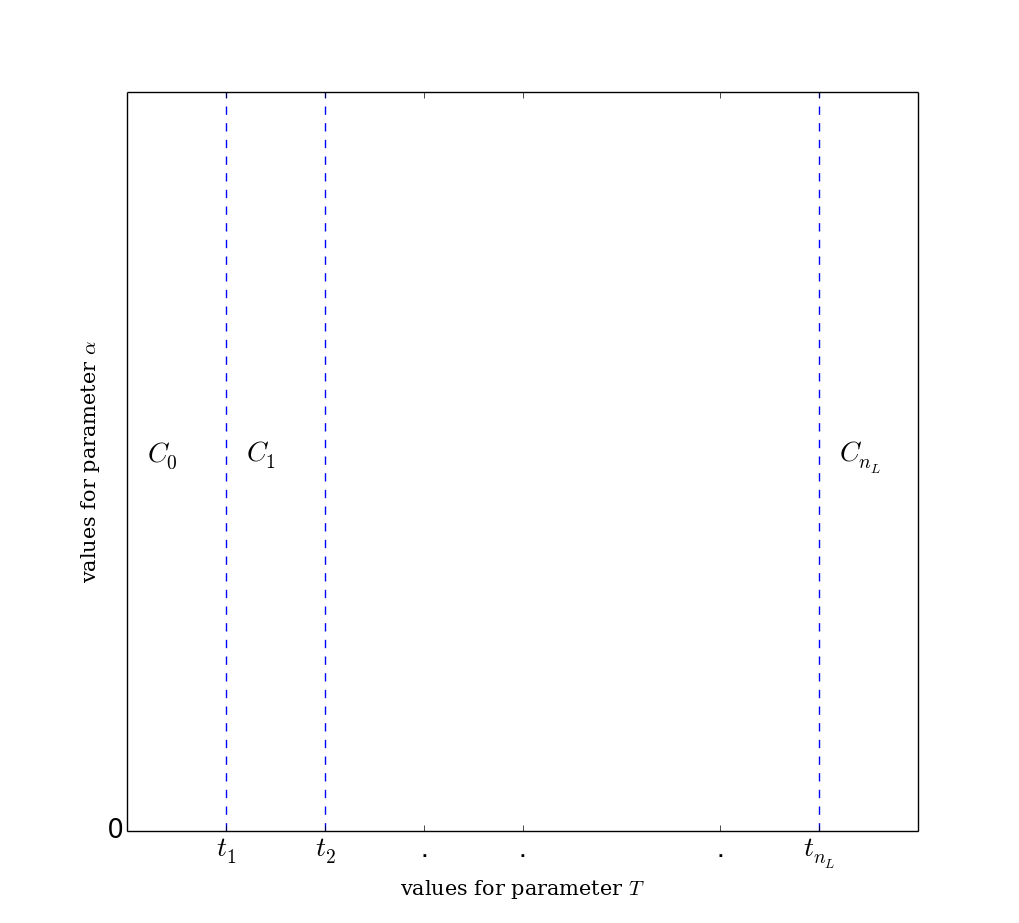}
  \caption{Cutting the parameter space, taking into account the discontinuities of the log-likelihood function in $\mathbb{R}^2$}
  \label{fig:DiscontLlkBot}
\end{figure}

For $i \in \{0, 1, ...,n_L+1\}$, let $C_i=\{(\alpha,T)\in R^+ \times R^+, t_i<T<t_{i+1} \}$  \\
\noindent with $t_0=0$ and $t_{n_L+1}=+\infty$.

Now let be $C=\bigcup_{i=0}^{n_L}C_i$ (Figure \ref{fig:DiscontLlkBot}). We can see that $\mathbb{L}_{SSPSC}(\alpha, T)$ is 
continuously differentiable in the interior of $C$.

Given that we do not consider negative values for $\alpha$ or $T$, we split the parameter space into two subsets $C$ and $\mathbb{R}^{2}_{+}\setminus C$.
If $(\alpha, T)\in C$, taking the log and the derivative with respect to $T$ in equation (\ref{eq:Likelihood_function}) gives:

\begin{equation*}
  \label{eq:partialT_LLk}
  \frac{\partial}{\partial T}\mathbb{L}_{t_i}(\alpha, T)=\begin{cases}
        -1+\frac{1}{\alpha} & \text{ if } T< t_i \\
        0 & \text{ otherwise.}
  \end{cases}
\end{equation*}

As we can see, if $\alpha<1$ then $\frac{\partial}{\partial T}\mathbb{L}_{t_i}(\alpha, T)>0$ 
for all $i$ and if $\alpha>1$ then $\frac{\partial}{\partial T}\mathbb{L}_{t_i}(\alpha, T)<0$ 
for all $i$. A consequence, $\nabla \mathbb{L}_{SSPSC}(\alpha, T)$ will never be zero in the interior of $C$ if 
$\alpha \neq 1$.
$\square$
\end{proof}

This fact suggests that the $min$ and $max$ values of $\mathbb{L}_{SSPSC}$ (if they exist) have the form $(\alpha, t_i)$.

Let's find the critical points of $\mathbb{L}_{SSPSC}$ over the lines $(\alpha, t_i)$ with 
$t_i \in \{t_1, t_2, ..., t_{n_L}\}$.  
When we fix the value of $T$ the function becomes a function of the single variable $\alpha$. If $T=t_a$ for $a\in \{1,...,n_L\}$
it follows from (\ref{eq:log-Likelihood_function}) that:

\begin{equation*}
  \label{eq:Llk_fixed_alpha}\\
  \log(\mathbb{L}_{SSPSC}(\alpha, t_a)) = \log\left(\frac{1}{\alpha}\right)-t_a-\displaystyle\sum_{i=1}^{n_L}\left[\log\left(\frac{1}{\alpha}\right)-t_a-\frac{1}{\alpha}(t_i-t_a)\right]\mathbb{I}_{t_a<t_i}-\displaystyle\sum_{i=1}^{n_L}t_i\mathbb{I}_{t_a>t_i}.
\end{equation*}

Denoting $K=\displaystyle\sum_{i=1}^{n_L}\mathbb{I}_{t_i < t_a}$, we then have:

\begin{equation*}
  \log(\mathbb{L}_{SSPSC}(\alpha, t_a)) = (K+1)\left(\log\left(\frac{1}{\alpha}\right)-t_a\right)-\frac{1}{\alpha}\displaystyle\sum_{i=1}^{n_L}(t_i-t_a)\mathbb{I}_{t_a<t_i}-\displaystyle\sum_{i=1}^{n_L}t_i\mathbb{I}_{t_a>t_i}.
\end{equation*}

Let us find the zeros of the derivative in $\alpha$:

\begin{equation*}
  \begin{split}
  \log(\mathbb{L}_{SSPSC}(\alpha, t_a)) = 0 
  \Leftrightarrow & -(K+1)\frac{1}{\alpha}+\frac{1}{\alpha^2}\displaystyle\sum_{i=1}^{n_L}(t_i-t_a)\mathbb{I}_{t_a<t_i} = 0 \\
\Leftrightarrow & -\alpha(K+1)+\displaystyle\sum_{i=1}^{n_L}(t_i-t_a)\mathbb{I}_{t_a<t_i} = 0 \\
\Leftrightarrow & \alpha = \frac{\displaystyle\sum_{i=1}^{n_L}t_i\mathbb{I}_{t_a<t_i} - Kt_a}{K+1}.
  \end{split}
\end{equation*}

Hence, the maximum value of the log-likelihood function (if it exists) is of the form: 

\begin{equation*}
  \label{eq:CritPoints}
  \begin{split}
    m_a=\left(\frac{\displaystyle\sum_{i=1}^{n_L}t_i\mathbb{I}_{t_a<t_i} - Kt_a}{K+1}, t_a\right) &, a \in \{1, 2, ..., n_L\}.
  \end{split}
\end{equation*}

We then take $\displaystyle (\hat{\alpha}, \hat{T}) = argmax_{a\in\{1, ..., n_L\}}\{\log(\mathbb{L}_{SSPSC}(m_a))\}$ as the Maximum Likelihood Estimation. 

\paragraph{Remark:}
Note that the function $\mathbb{L}_{SSPSC}$ does not have an upper bound. 
Let's say $(T_1, T_2, ..., T_{n_L})$ are the $n_L$ observations of $T_2$ sorted form the lower value to the higher value.

For $T=T_{n_L}$ we have from (\ref{eq:log-Likelihood_function}) and (\ref{eq:Likelihood_function}):

\begin{equation*}
 \mathbb{L}_{SSPSC}(\alpha, T_{n_L}) = log(\frac{1}{\alpha}) - T_{n_L} - \displaystyle\sum_{i=1}^{n_L-1}T_i
\end{equation*}

\noindent which clearly goes to $+\infty$ as $\alpha$ goes to zero.
The same problem is present for every $T\in ]T_{n_L-1}, T_{n_L}]$

Nevertheless, in practice, our $m_a$ is a good estimate for the parameters.

\clearpage

\section{Proportion of rejections}

\begin{figure}[h]
  \centering
  \includegraphics[width=\textwidth,type=png,ext=.png,read=.png]{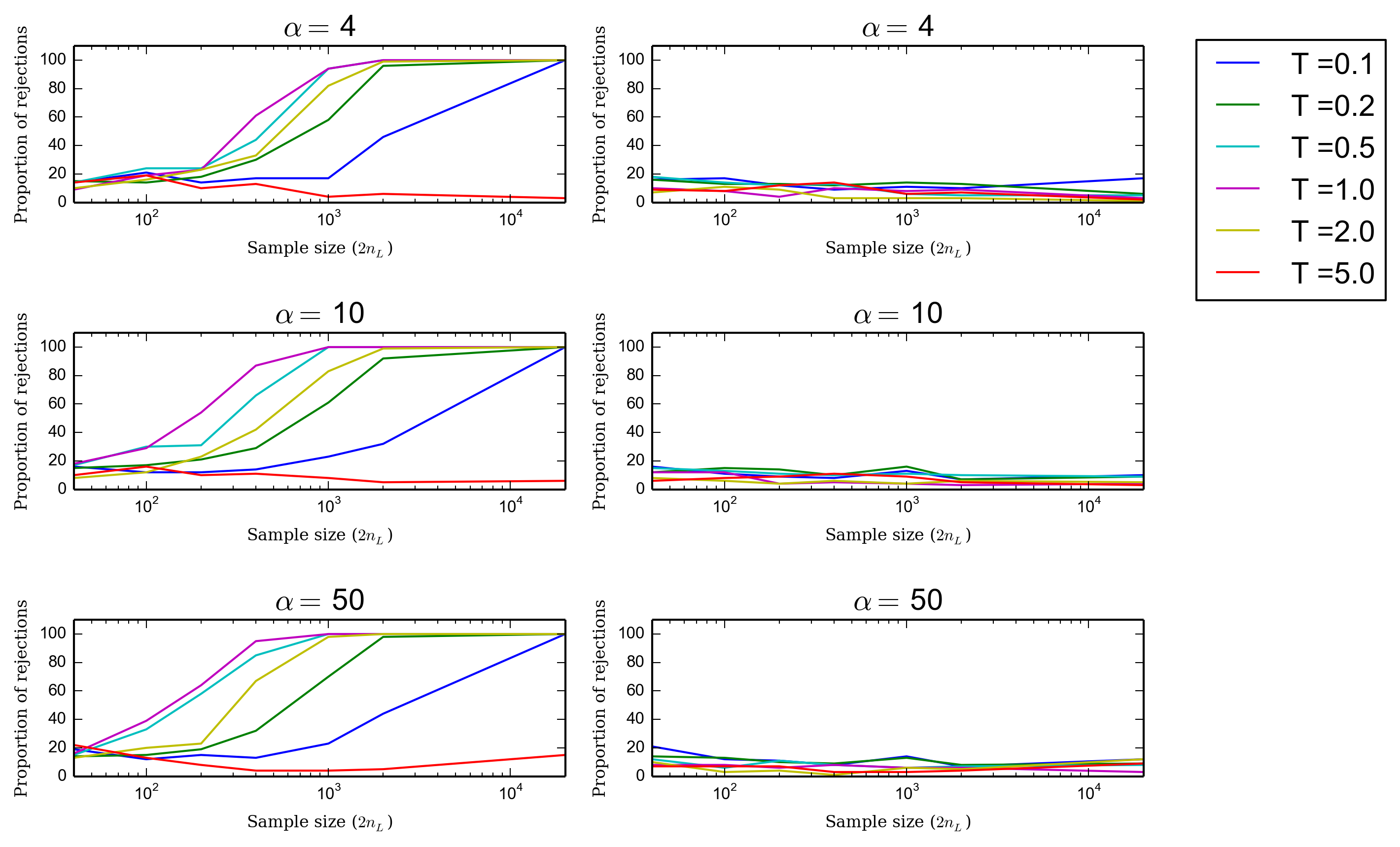}
  \caption{Proportion of rejected data sets simulated under the SSPSC model. Left panels: the reference model is the StSI model (the wrong model). Right panels: the reference model is the SSPSC (the right model), \textit{i.e.} the model under which the data were simulated. Note that for the abscissa we used $2n_L$ instead of $n_L$ because in order to perform the $KS$ test it is necessary to first estimate the parameters using $n_L$ loci and then an independent set of $n_L$ values of $T_2$.}
\end{figure}

\begin{figure}[h]
  \centering
  \includegraphics[width=\textwidth,type=png,ext=.png,read=.png]{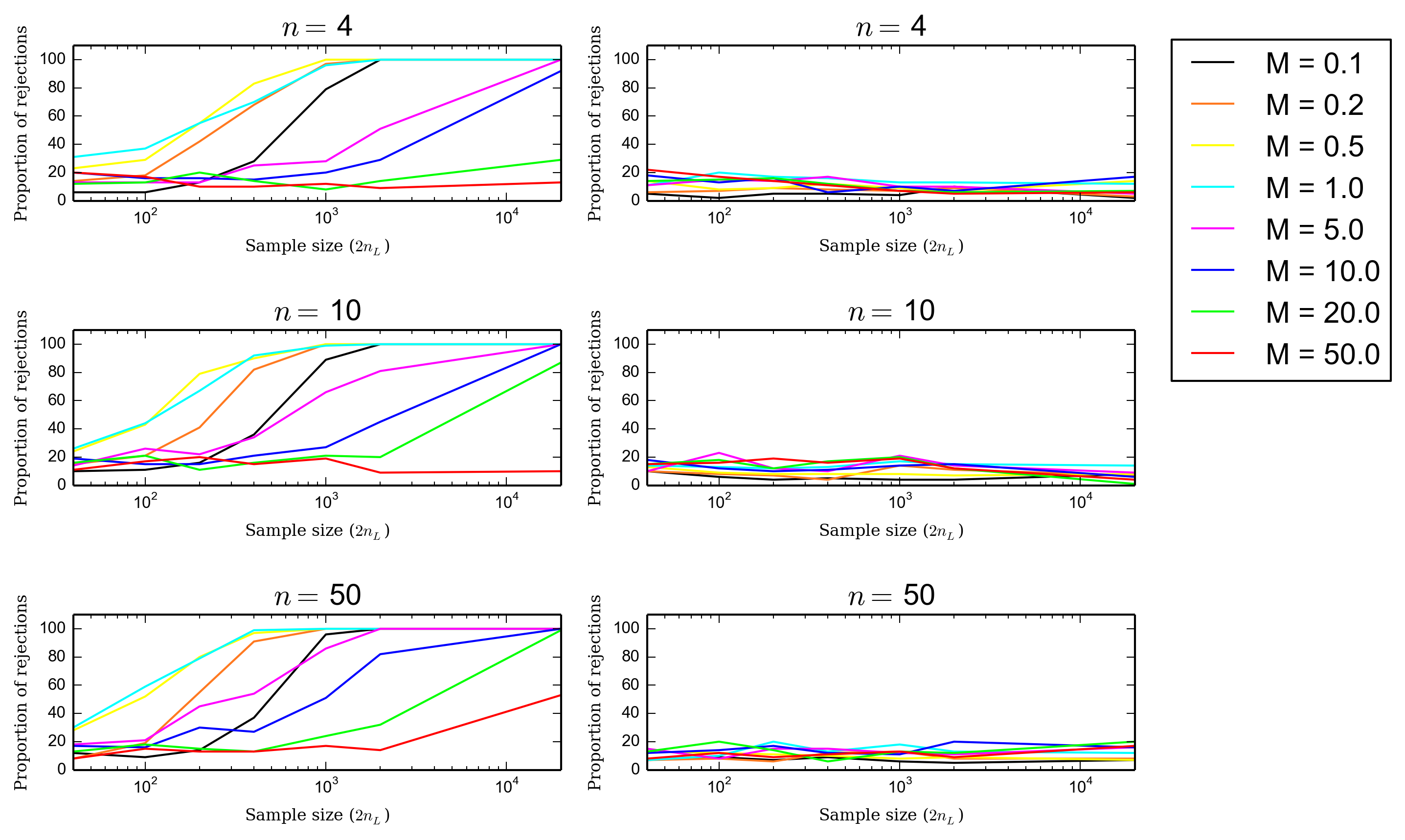}
  \caption{Proportion of rejected data sets simulated under the StSI model. Left panels: the reference model is the SSPSC (the wrong model). Right panels: the reference model is the StSI model (the right model), \textit{i.e.} the model under which the data were simulated. Note that for the abscissa we used $2n_L$ instead of $n_L$ because in order to perform the $KS$ test it is necessary to first estimate the parameters using $n_L$ loci and then an independent set of $n_L$ values of $T_2$.}
\end{figure}

\clearpage

\section{Estimation accuracy}

\begin{figure}[h]
  \centering
  \includegraphics[width=\textwidth,type=png,ext=.png,read=.png]{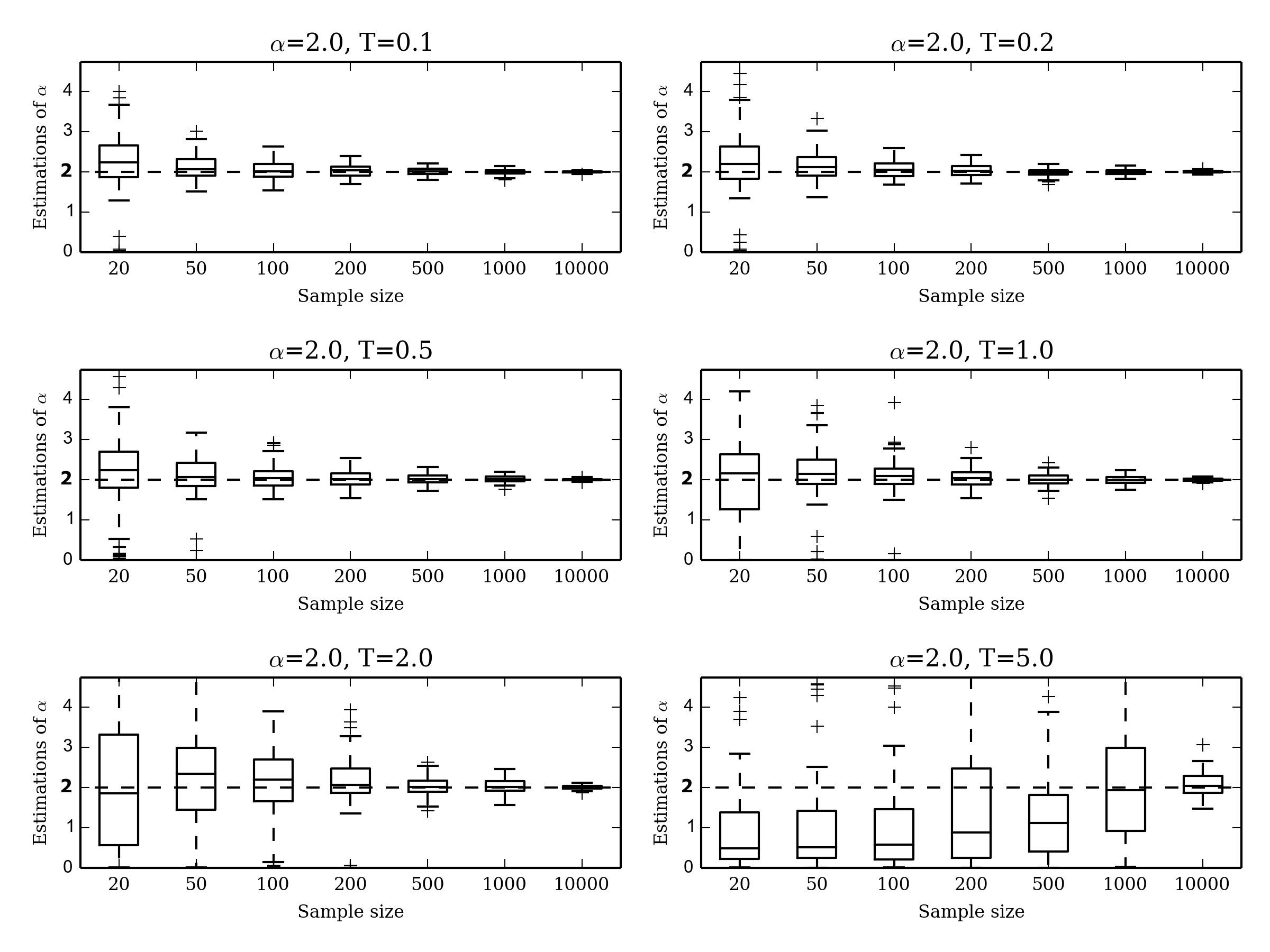}
  \caption{Estimations of $\alpha$ (real value $\alpha=2.0$)}
  \label{fig:alpha_2}
\end{figure}

\begin{figure}[h]
  \centering
  \includegraphics[width=\textwidth,type=png,ext=.png,read=.png]{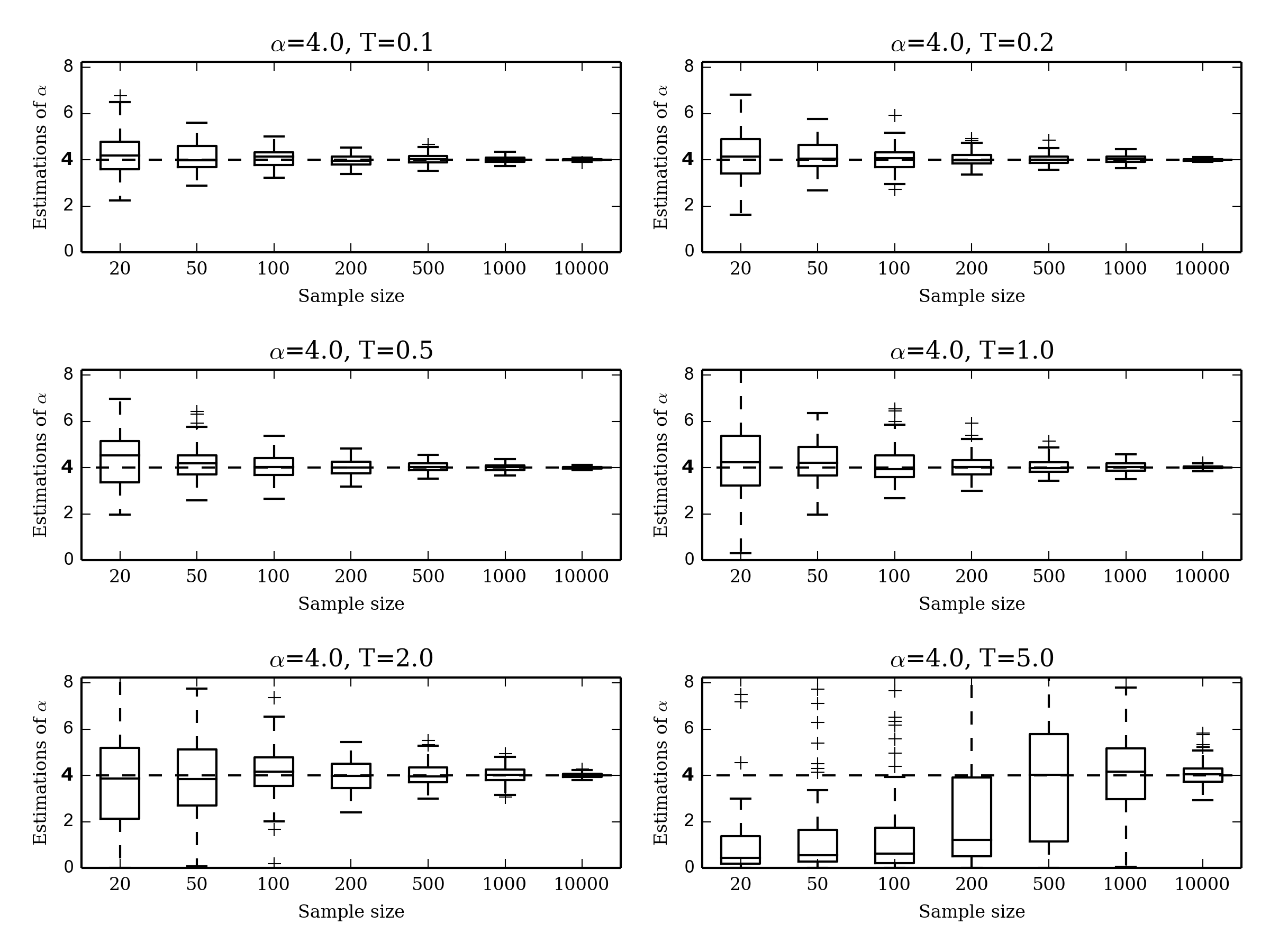}
  \caption{Estimations of $\alpha$ (real value $\alpha=4.0$)}
  \label{fig:alpha_4}
\end{figure}

\begin{figure}[h]
  \centering
  \includegraphics[width=\textwidth,type=png,ext=.png,read=.png]{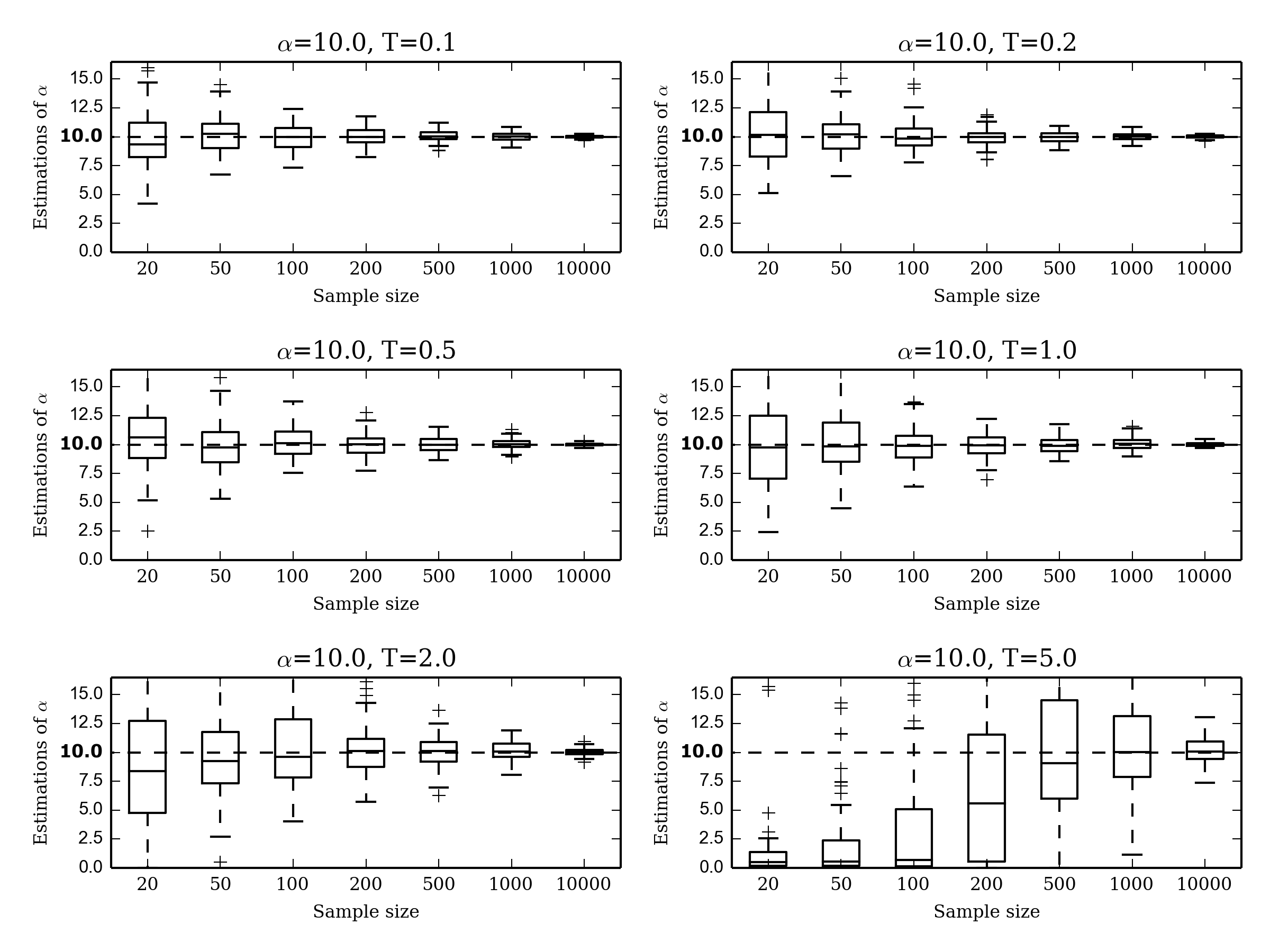}
  \caption{Estimations of $\alpha$ (real value $\alpha=10.0$)}
  \label{fig:alpha_10}
\end{figure}

\begin{figure}[h]
  \centering
  \includegraphics[width=\textwidth,type=png,ext=.png,read=.png]{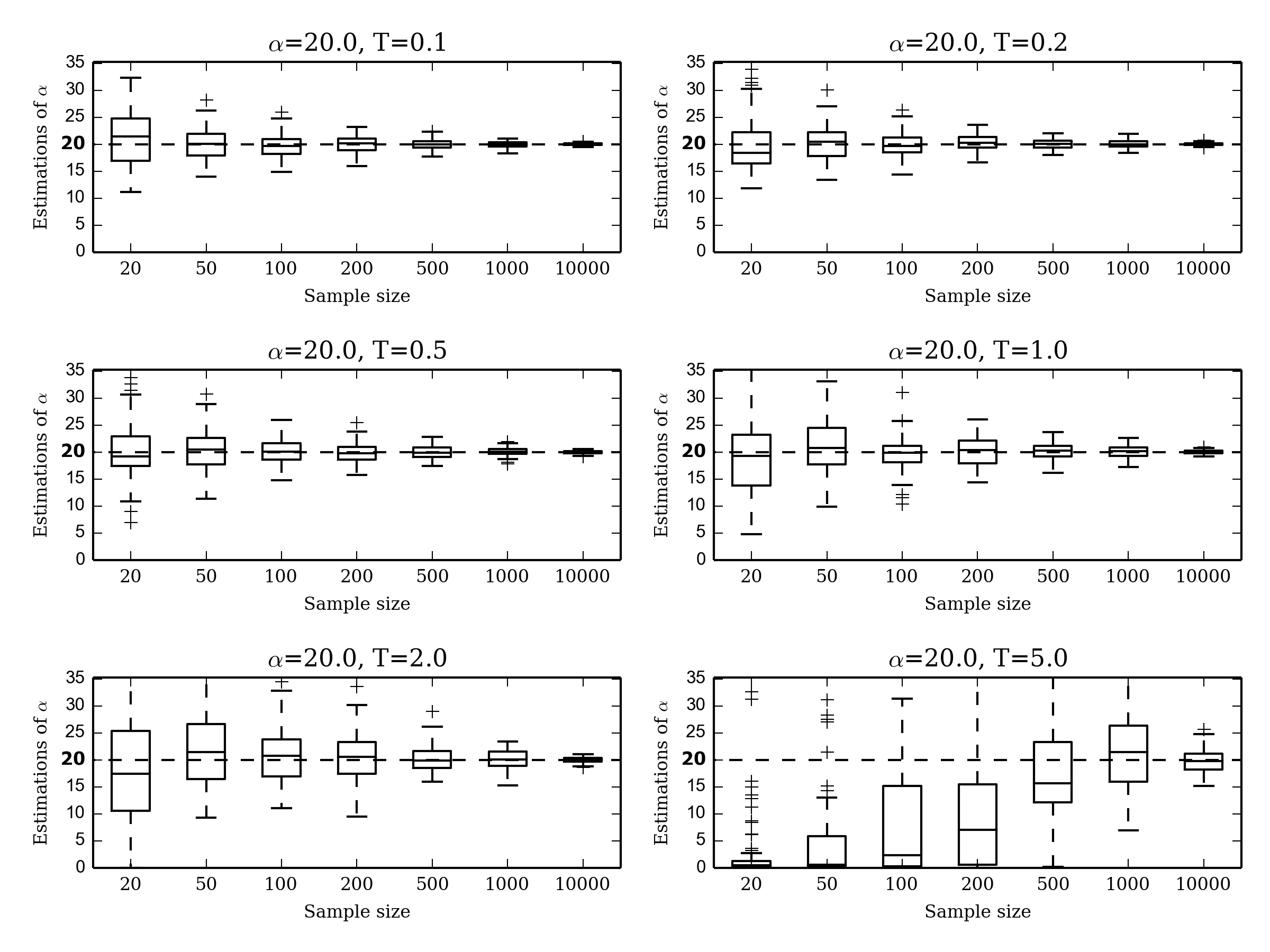}
  \caption{Estimations of $\alpha$ (real value $\alpha=20.0$)}
  \label{fig:alpha_20}
\end{figure}

\begin{figure}[h]
  \centering
  \includegraphics[width=\textwidth,type=png,ext=.png,read=.png]{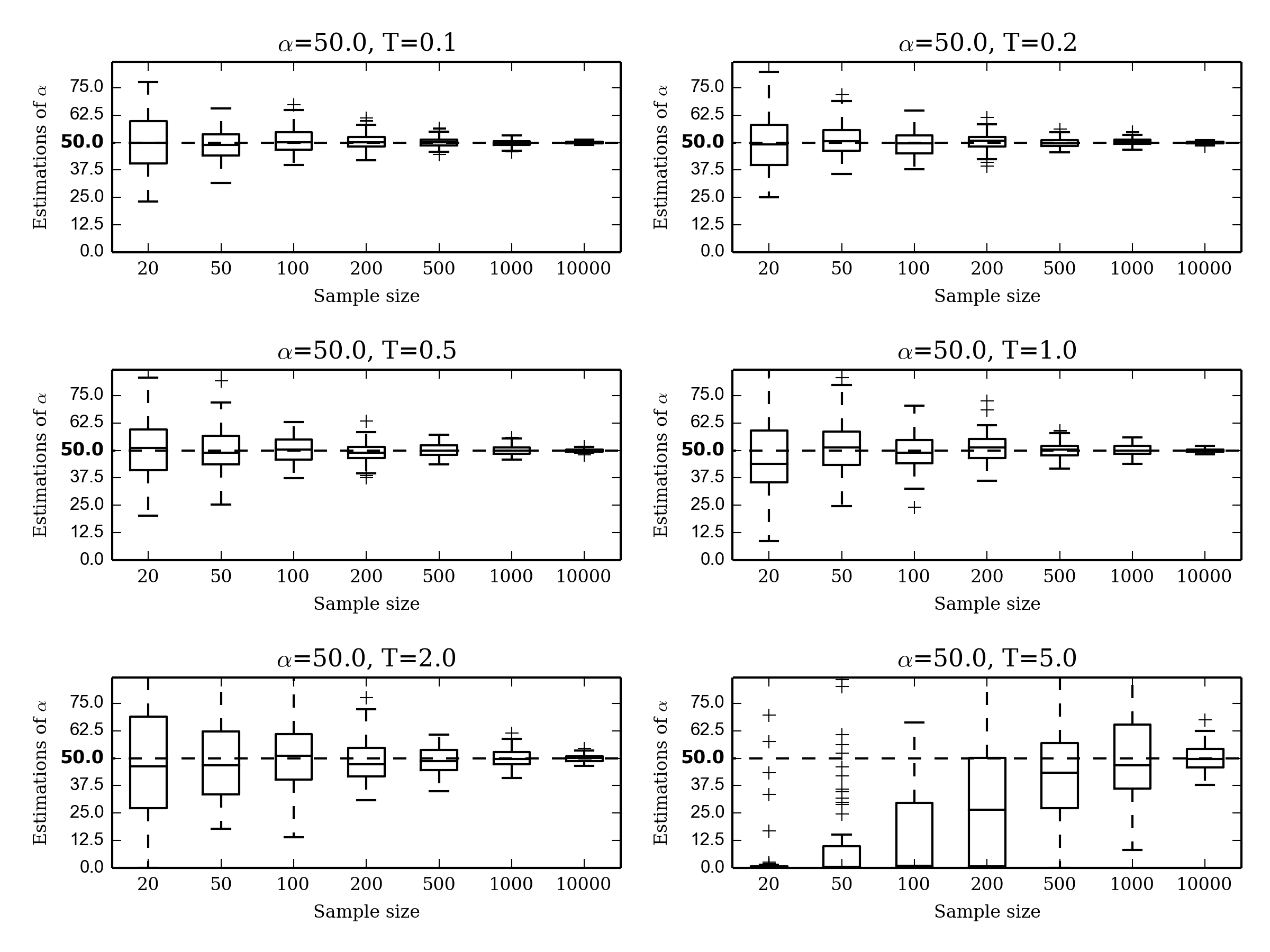}
  \caption{Estimations of $\alpha$ (real value $\alpha=50.0$)}
  \label{fig:alpha_50}
\end{figure}

\begin{figure}[h]
  \centering
  \includegraphics[width=\textwidth,type=png,ext=.png,read=.png]{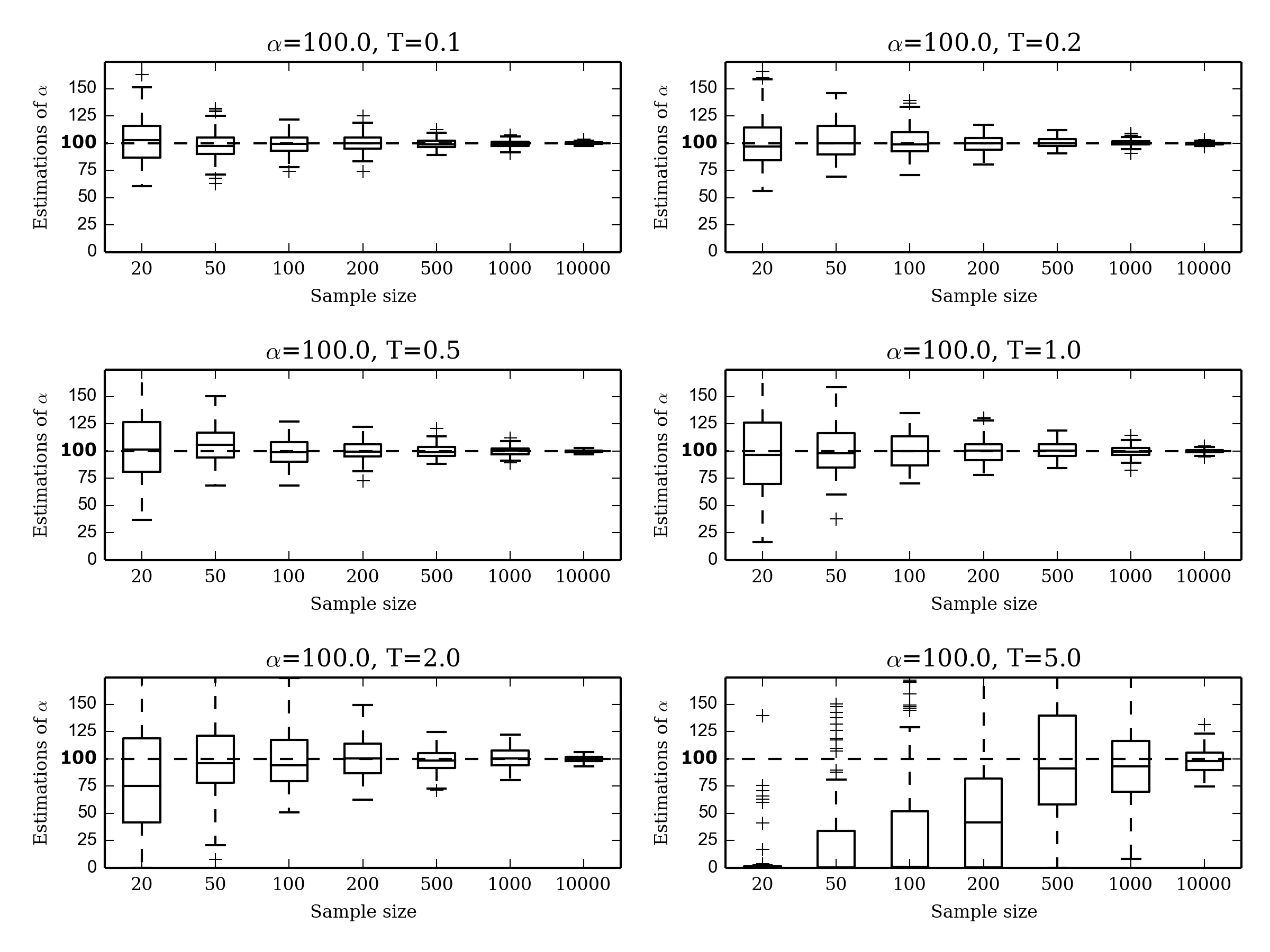}
  \caption{Estimations of $\alpha$ (real value $\alpha=100.0$)}
  \label{fig:alpha_100}
\end{figure}

All the scripts used in the simulations are available from \url{http://github.com/willyrv/SSPSC_vs_StSI} \\
Figures \ref{fig:alpha_2}, \ref{fig:alpha_4}, \ref{fig:alpha_10}, \ref{fig:alpha_20}, \ref{fig:alpha_50} and \ref{fig:alpha_100} show the results of the estimation of $\alpha$ for all the values of $T$ and all the sample sizes, under the SSPSC model. The values where $\alpha=(2, 4, 10, 20, 50, 100)$, $T=(0.1, 0.2, 0.5, 1, 2, 5)$ and $n_L=(20, 50, 100, 200, 500, 1000, 10000)$

\begin{figure}[h]
  \centering
  \includegraphics[width=\textwidth,type=png,ext=.png,read=.png]{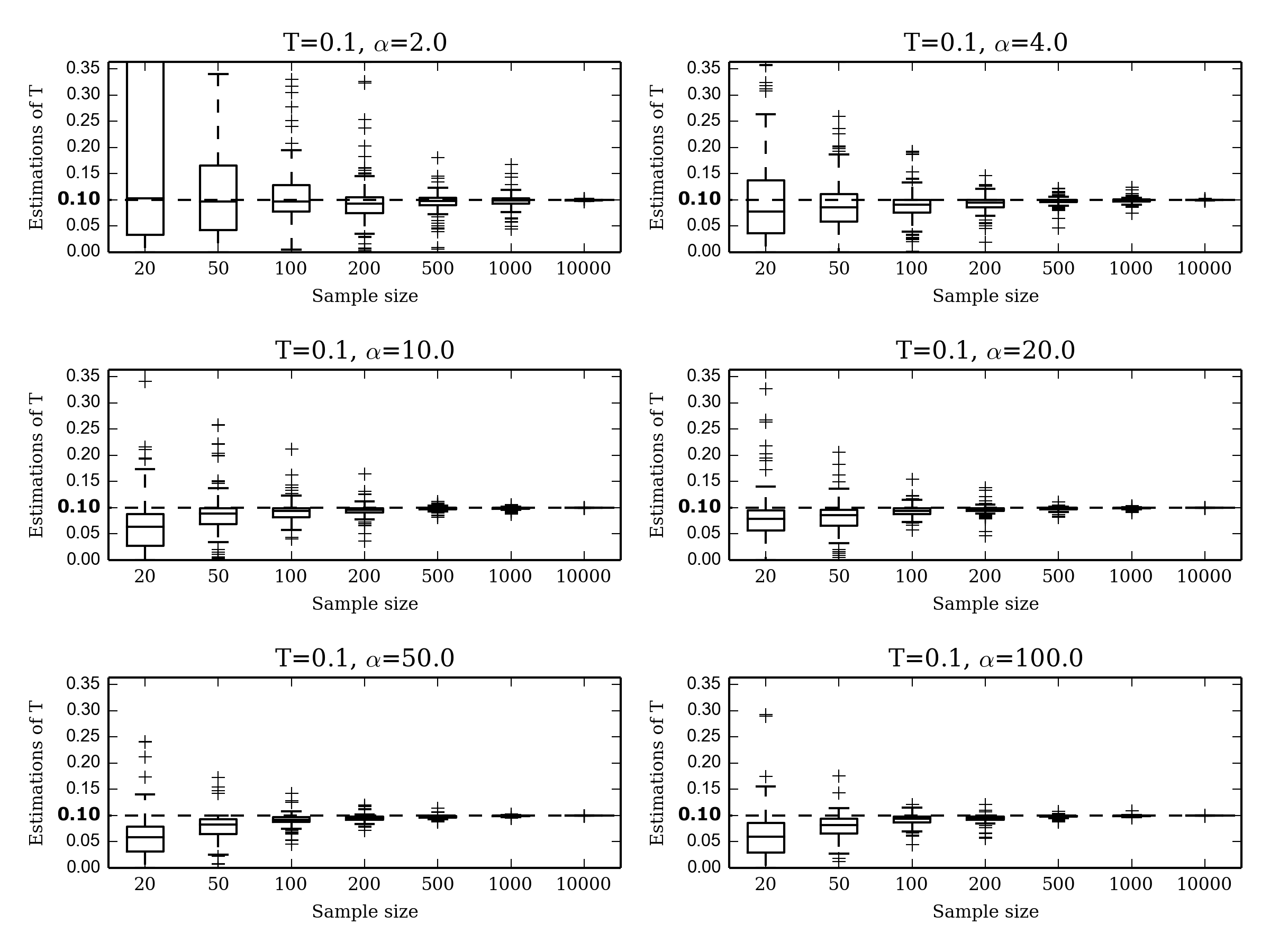}
  \caption{Estimations of $T$ (real value $T=0.1$)}
  \label{fig:T_0.1}
\end{figure}

\begin{figure}[h]
  \centering
  \includegraphics[width=\textwidth,type=png,ext=.png,read=.png]{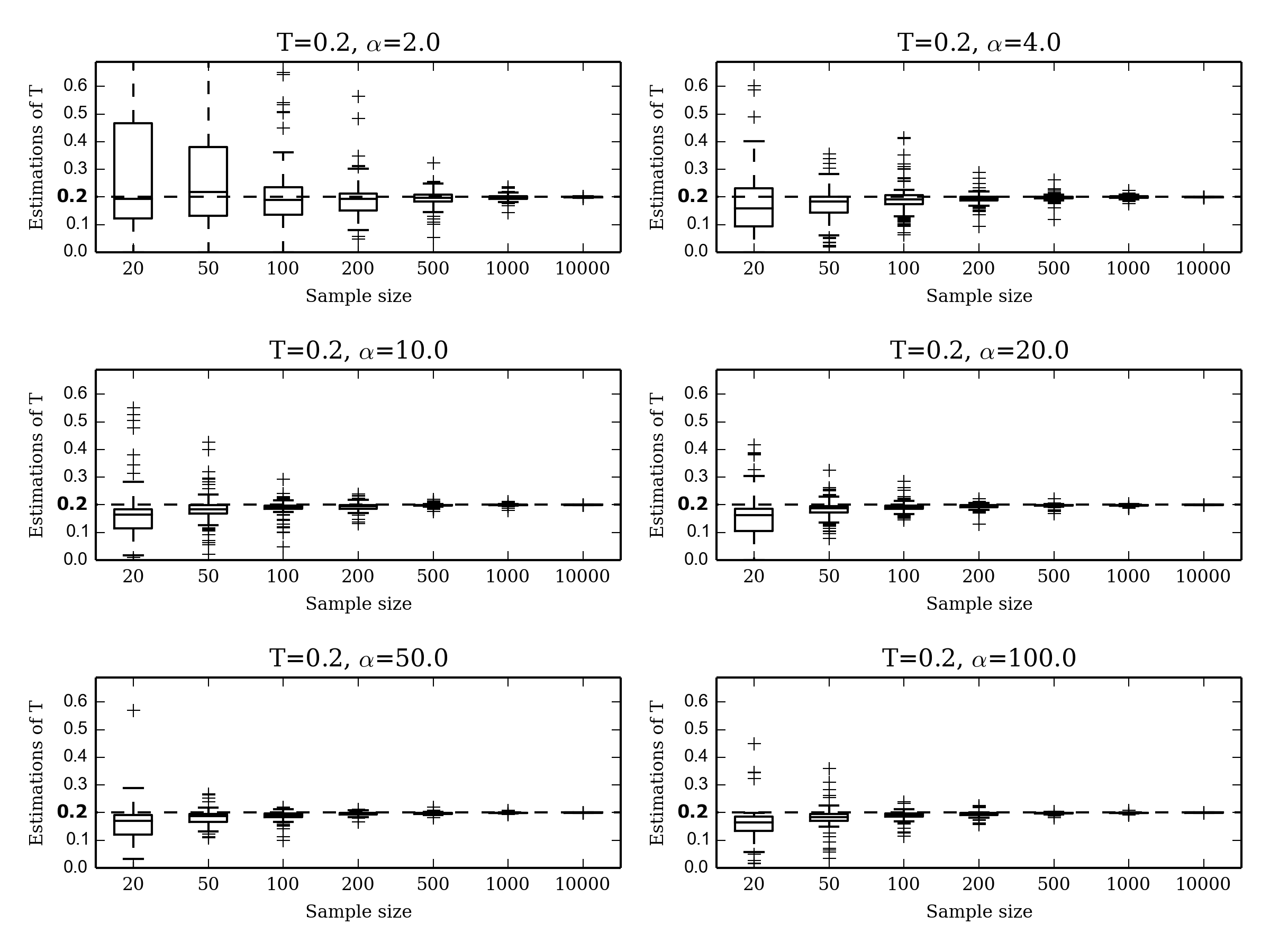}
  \caption{Estimations of $T$ (real value $T=0.2$)}
  \label{fig:T_02}
\end{figure}

\begin{figure}[h]
  \centering
  \includegraphics[width=\textwidth,type=png,ext=.png,read=.png]{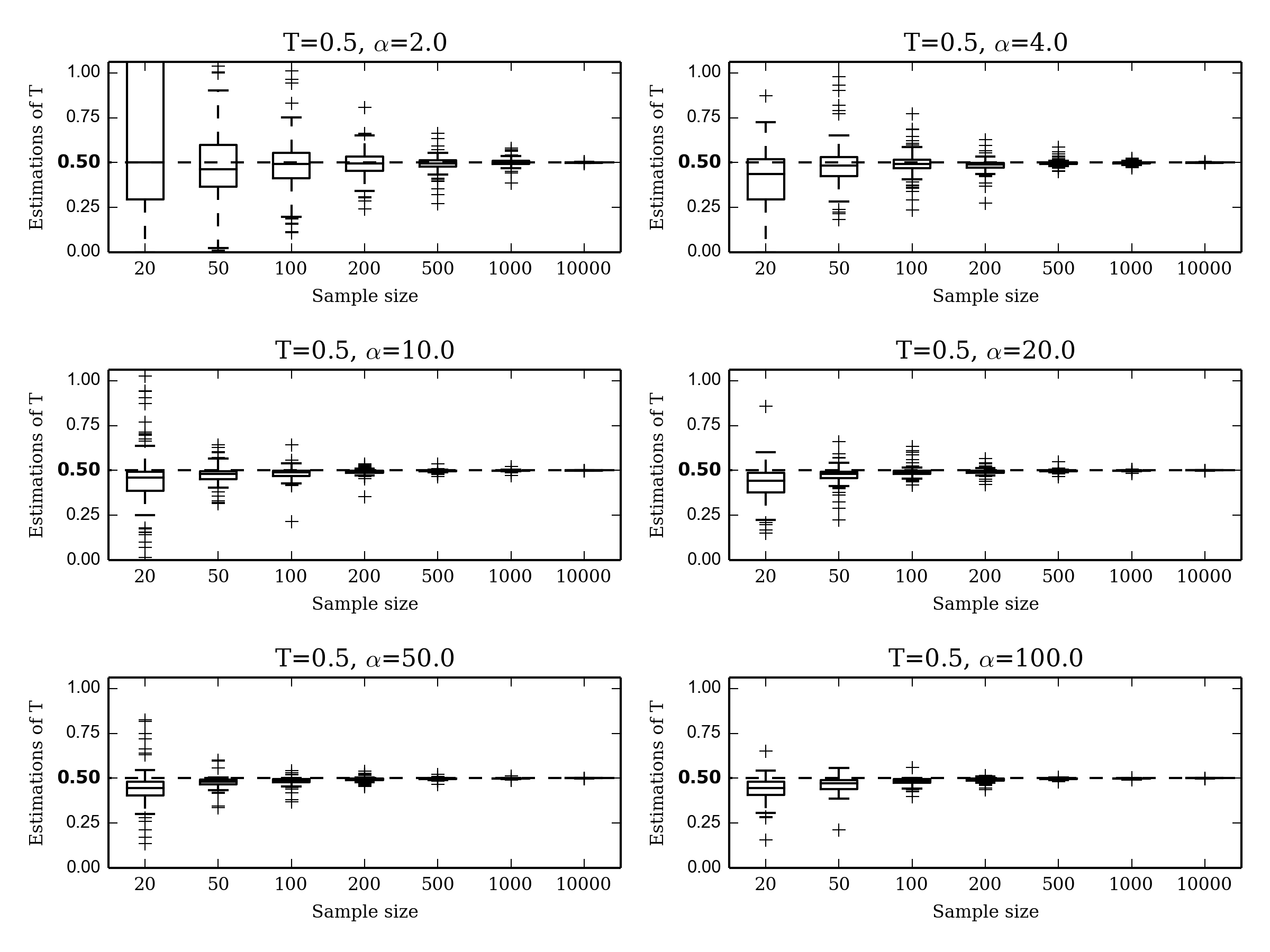}
  \caption{Estimations of $T$ (real value $T=0.5$)}
  \label{fig:T_0.5}
\end{figure}

\begin{figure}[h]
  \centering
  \includegraphics[width=\textwidth,type=png,ext=.png,read=.png]{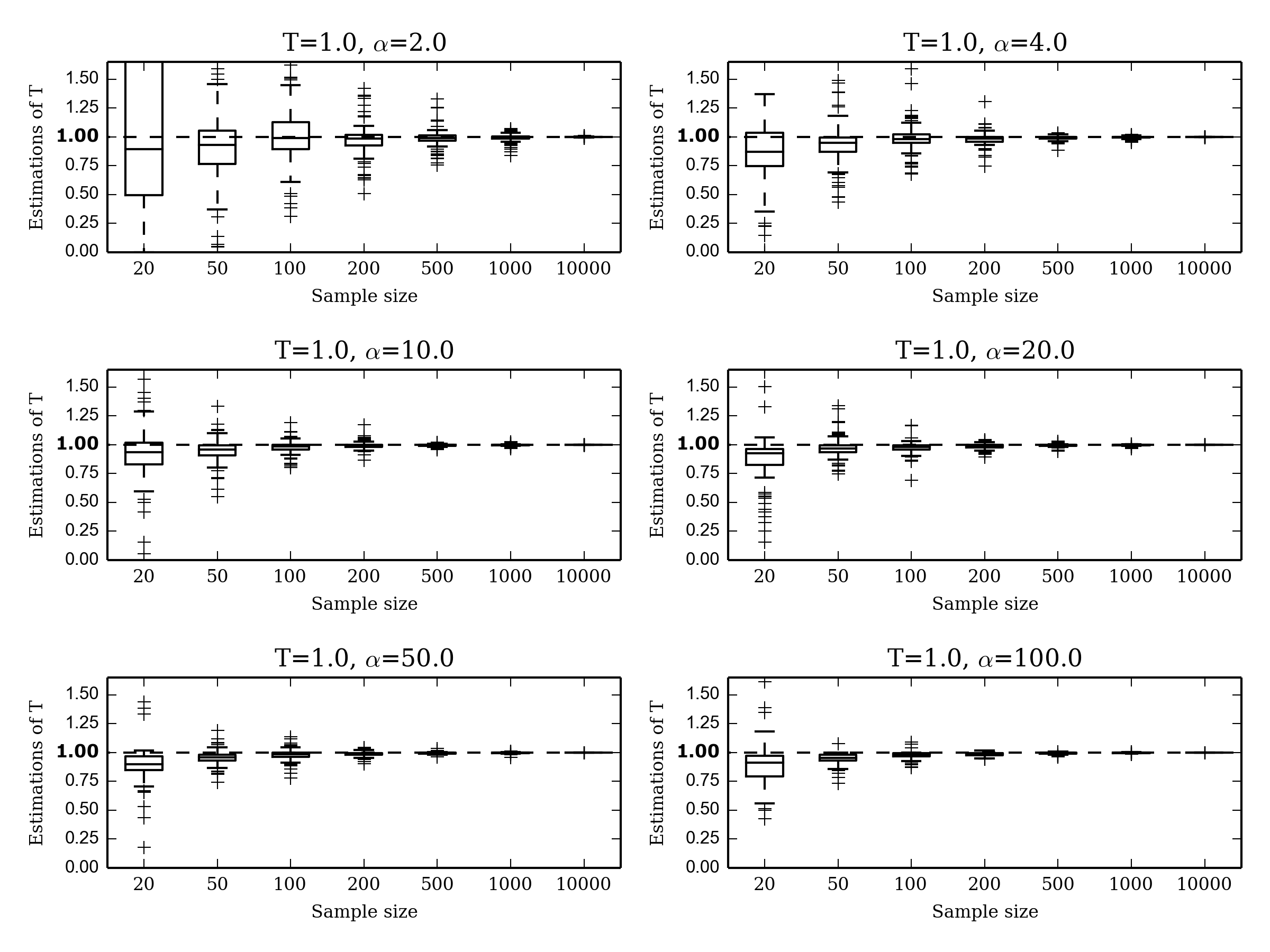}
  \caption{Estimations of $T$ (real value $T=1.0$)}
  \label{fig:T_1}
\end{figure}

\begin{figure}[h]
  \centering
  \includegraphics[width=\textwidth,type=png,ext=.png,read=.png]{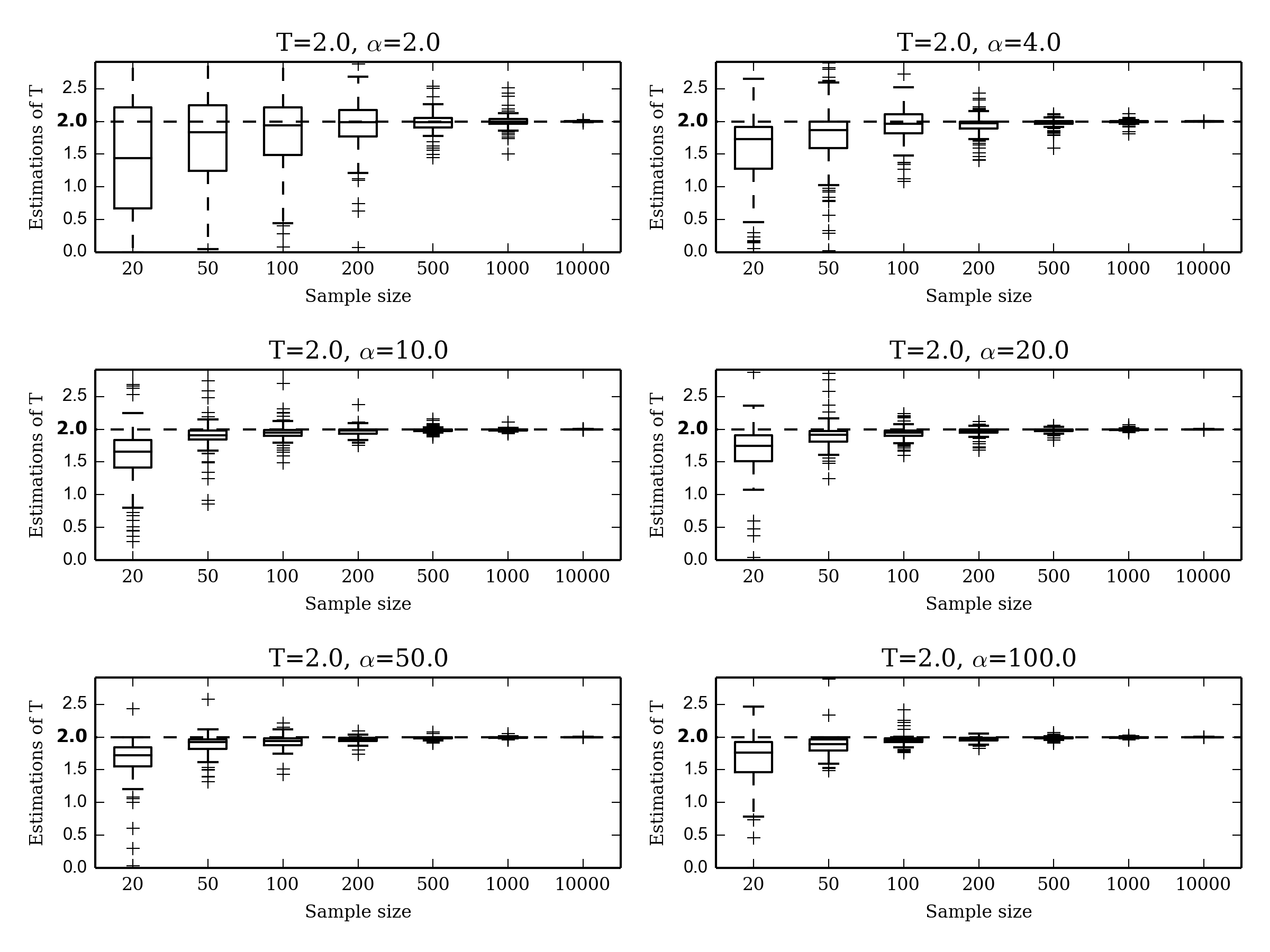}
  \caption{Estimations of $T$ (real value $T=2.0$)}
  \label{fig:T_2}
\end{figure}

\begin{figure}[h]
  \centering
  \includegraphics[width=\textwidth,type=png,ext=.png,read=.png]{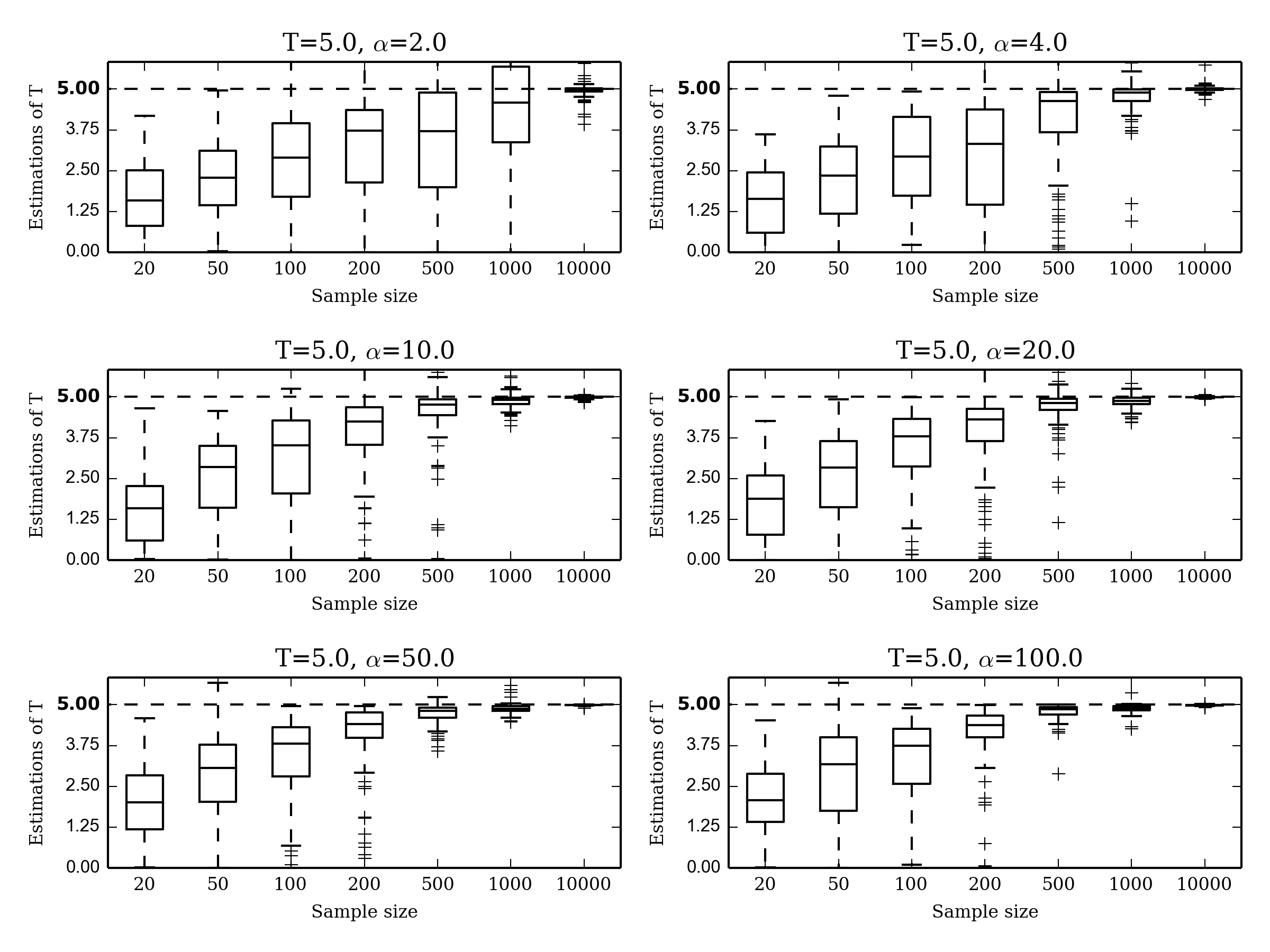}
  \caption{Estimations of $T$ (real value $T=5.0$)}
  \label{fig:T_5}
\end{figure}

\clearpage
Figures \ref{fig:T_0.1}, \ref{fig:T_02}, \ref{fig:T_0.5}, \ref{fig:T_1}, \ref{fig:T_2} and \ref{fig:T_5} show the results of the estimation of $T$ for all the values of $\alpha$ and all the sample sizes, under the SSPSC model. The values where $\alpha=(2, 4, 10, 20, 50, 100)$, $T=(0.1, 0.2, 0.5, 1, 2, 5)$ and $n_L=(20, 50, 100, 200, 500, 1000, 10000)$

\begin{figure}[h](0.1, 0.2, 0.5, 1, 5, 10, 20, 50) and (2, 4, 10, 20, 50, 100)
  \centering
  \includegraphics[width=\textwidth,type=png,ext=.png,read=.png]{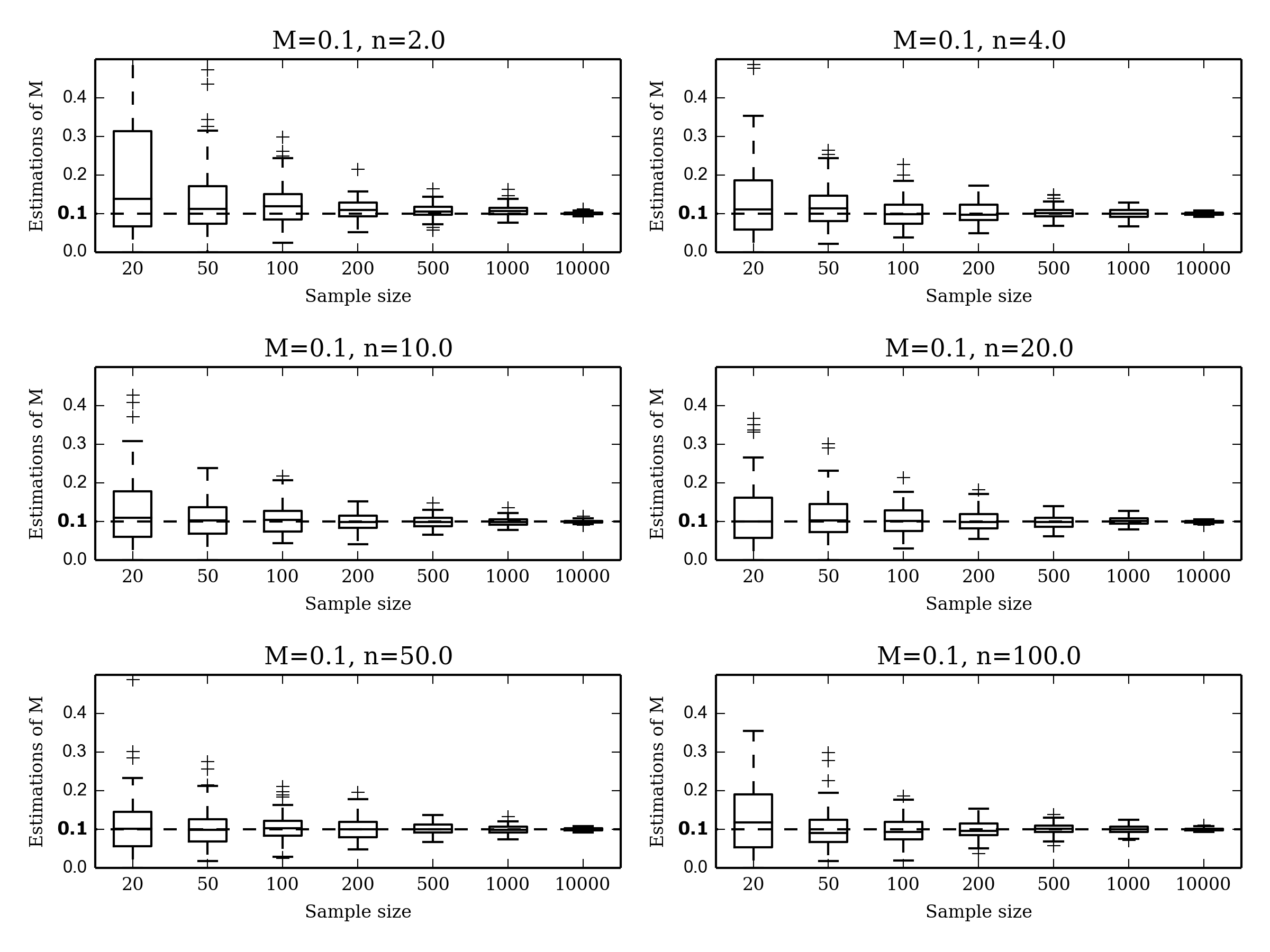}
  \caption{Estimations of $M$ (real value $M=0.1$)}
  \label{fig:M_0.1}
\end{figure}

\begin{figure}[h]
  \centering
  \includegraphics[width=\textwidth,type=png,ext=.png,read=.png]{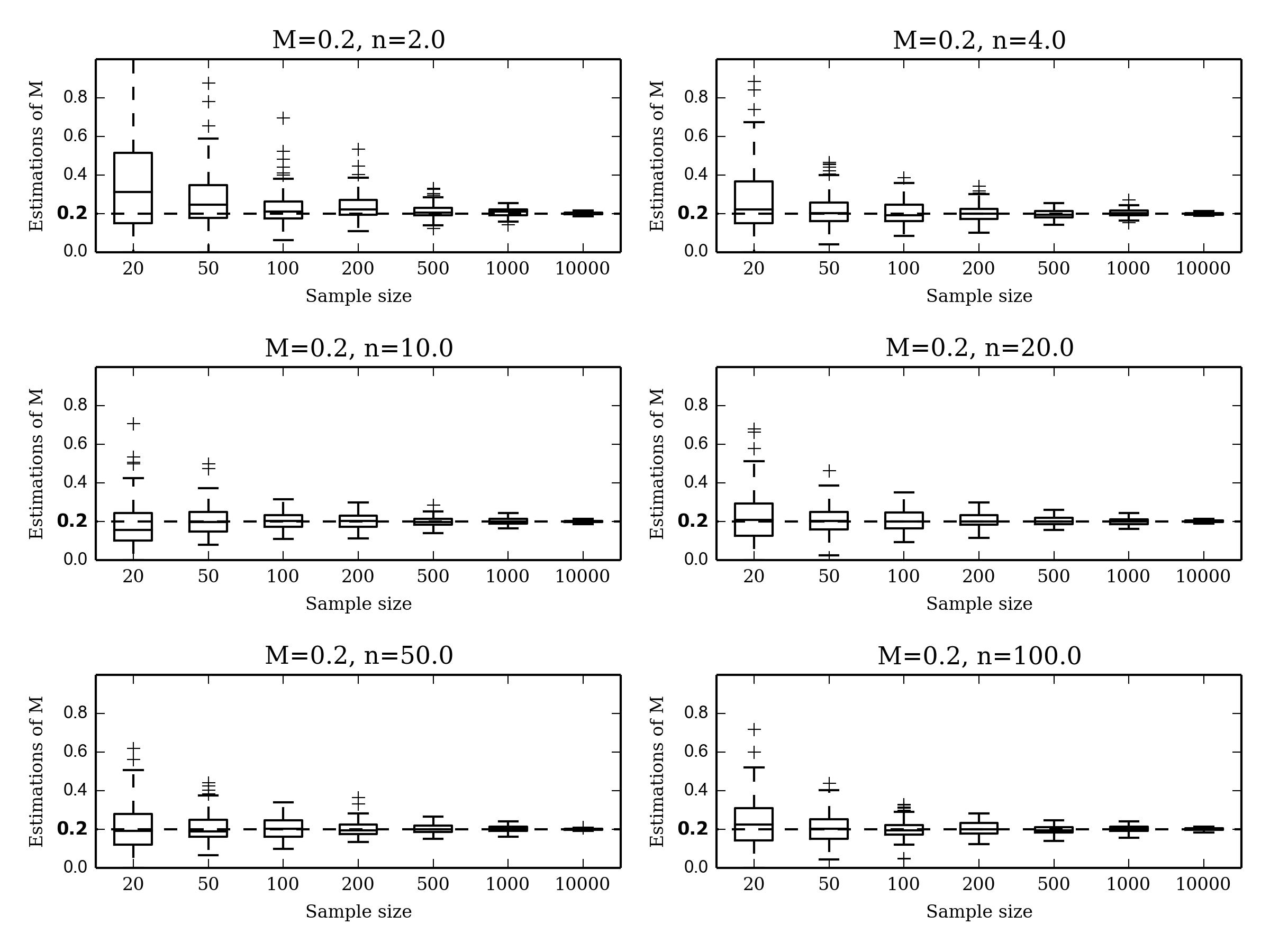}
  \caption{Estimations of $M$ (real value $M=0.2$)}
  \label{fig:M_0.2}
\end{figure}

\begin{figure}[h]
  \centering
  \includegraphics[width=\textwidth,type=png,ext=.png,read=.png]{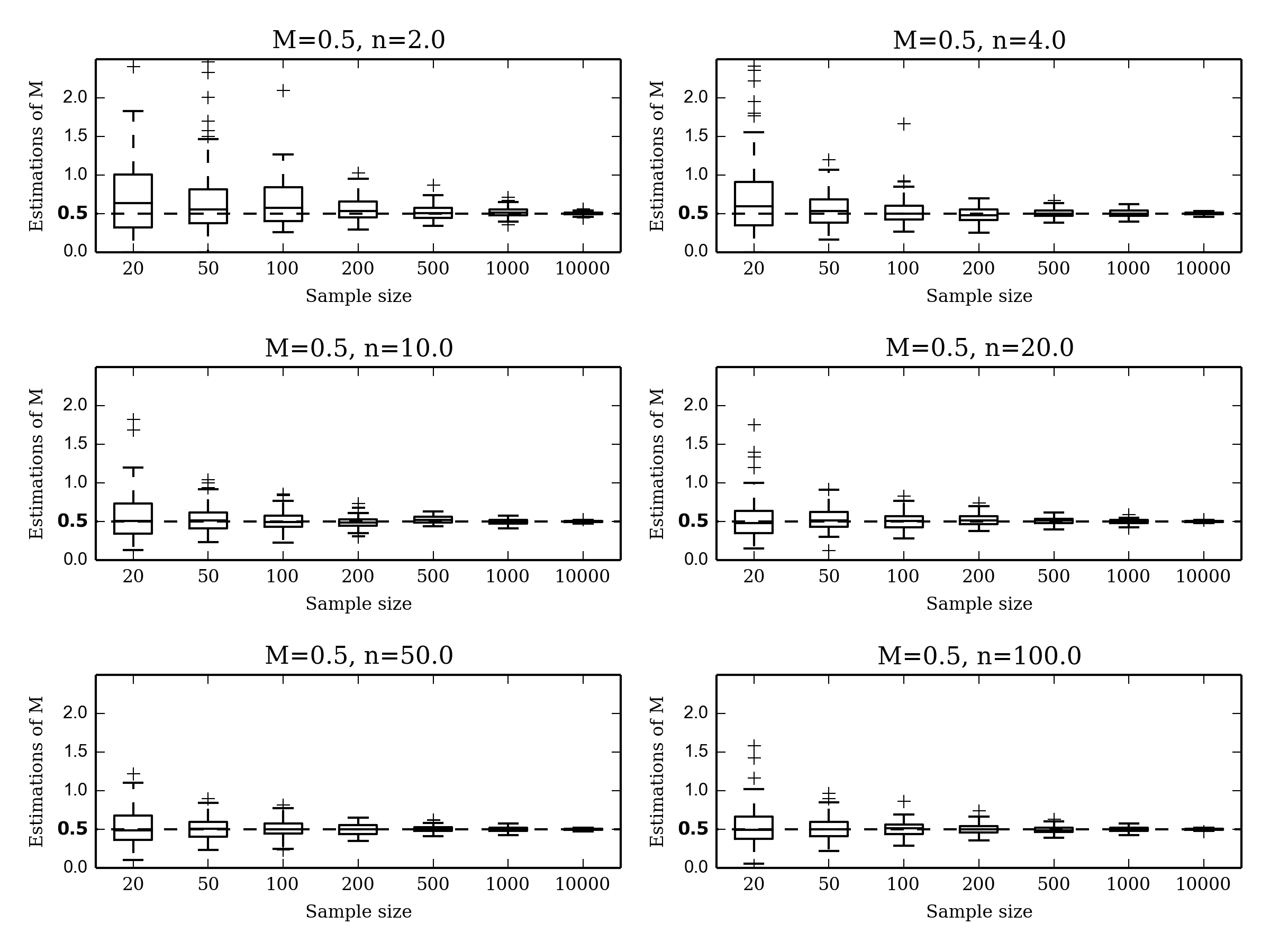}
  \caption{Estimations of $M$ (real value $M=0.5$)}
  \label{fig:M_0.5}
\end{figure}

\begin{figure}[h]
  \centering
  \includegraphics[width=\textwidth,type=png,ext=.png,read=.png]{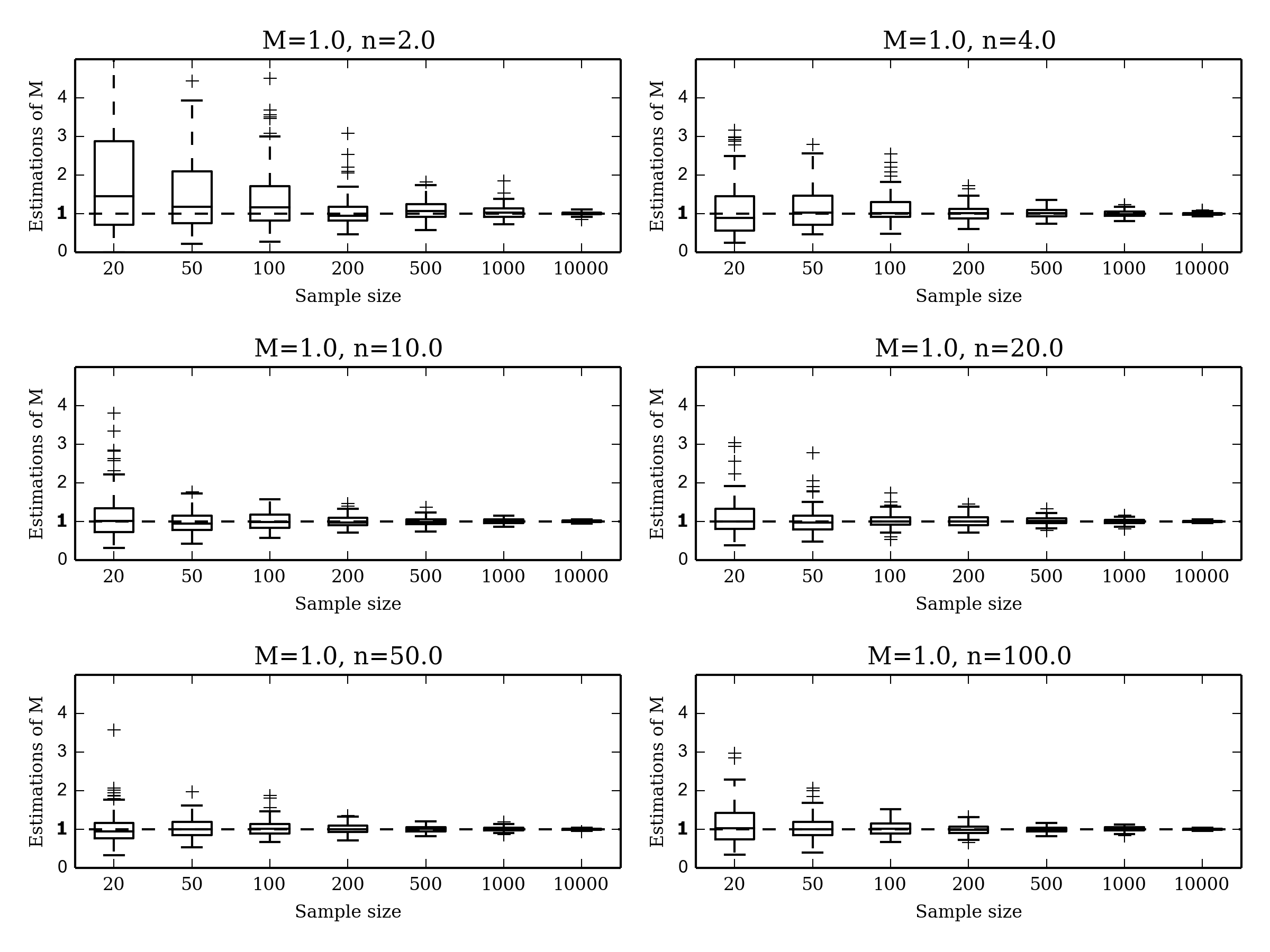}
  \caption{Estimations of $M$ (real value $M=1.0$)}
  \label{fig:M_1}
\end{figure}

\begin{figure}[h]
  \centering
  \includegraphics[width=\textwidth,type=png,ext=.png,read=.png]{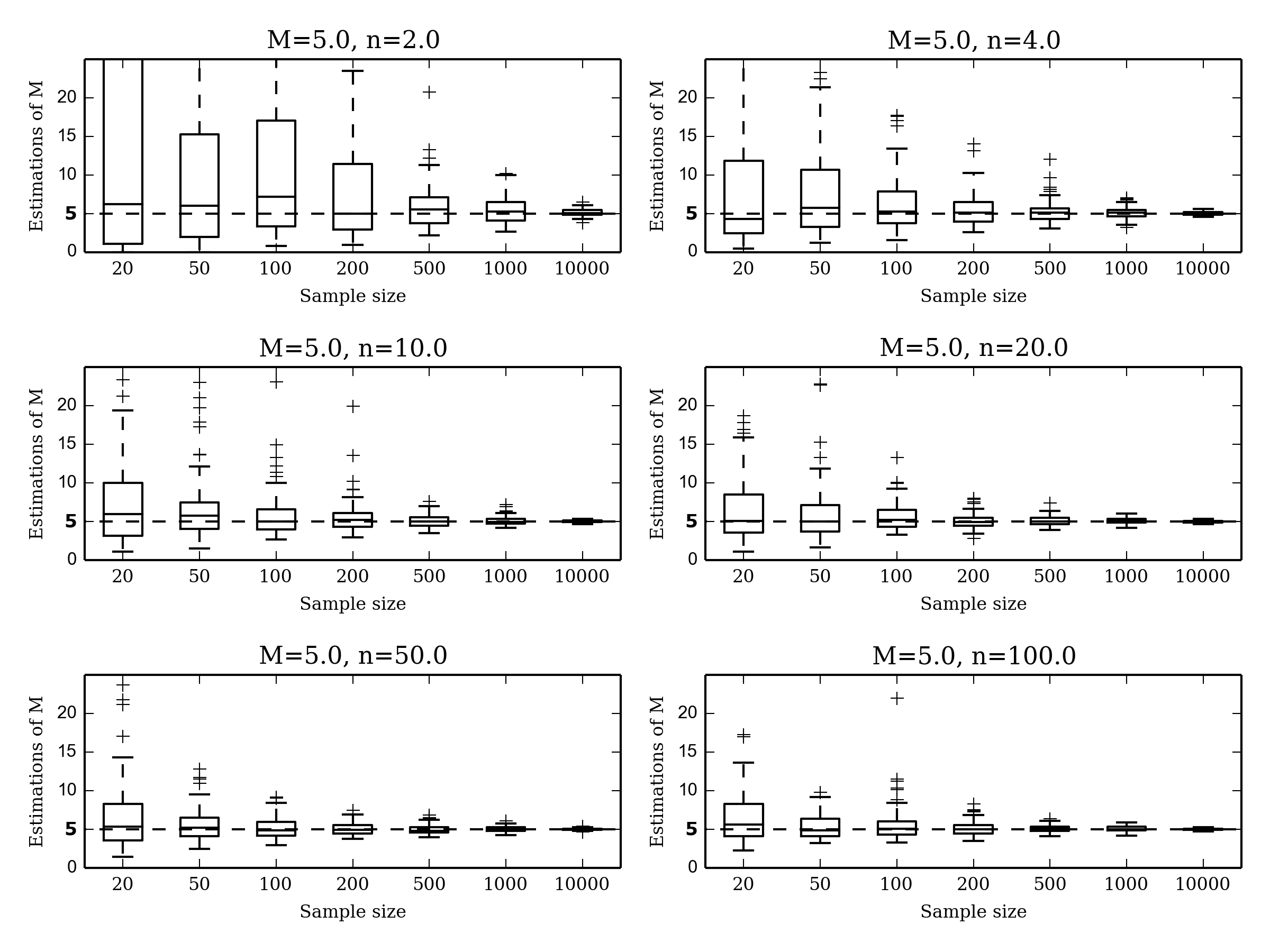}
  \caption{Estimations of $M$ (real value $M=5.0$)}
  \label{fig:M_5}
\end{figure}

\begin{figure}[h]
  \centering
  \includegraphics[width=\textwidth,type=png,ext=.png,read=.png]{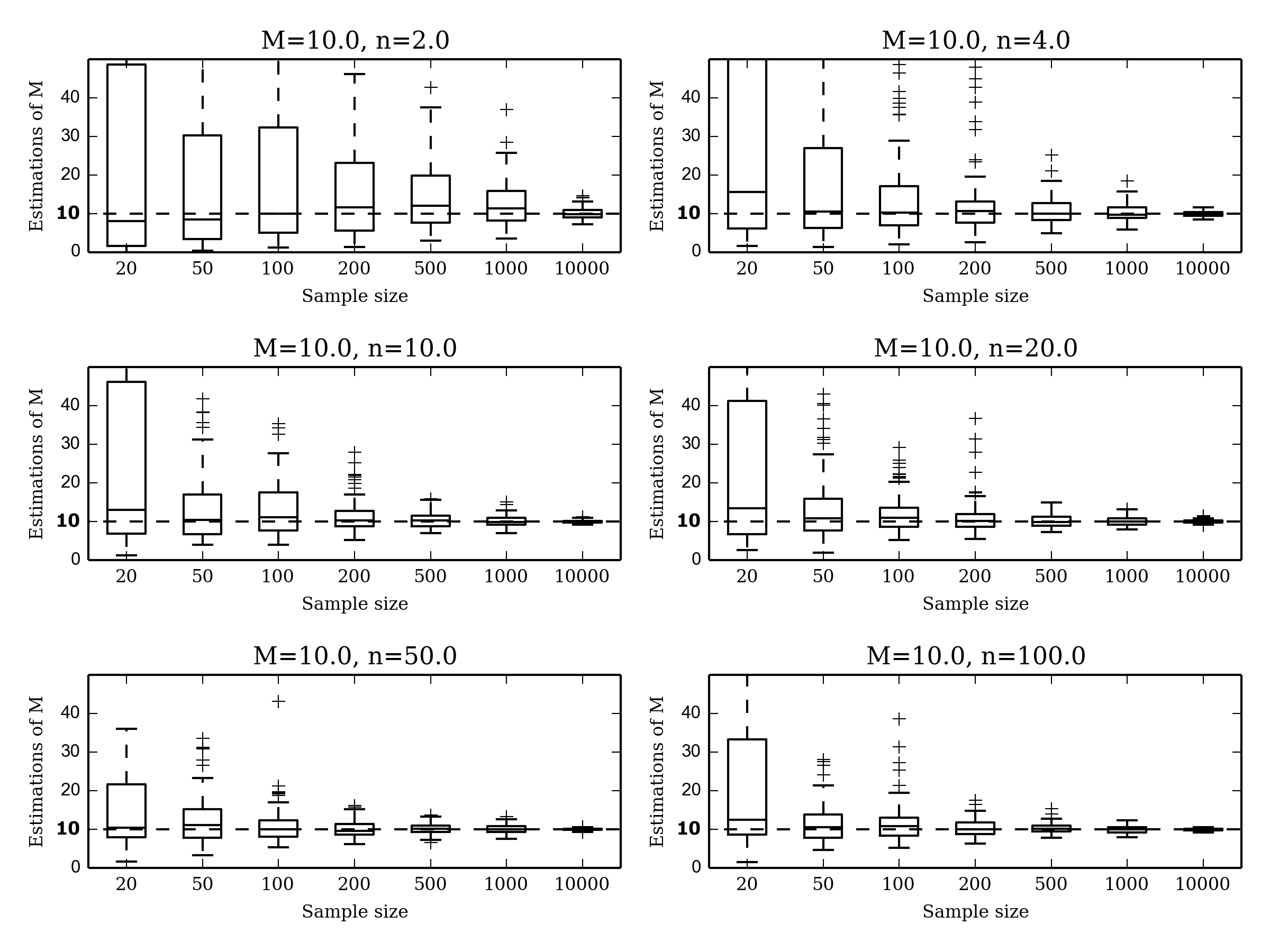}
  \caption{Estimations of $M$ (real value $M=10.0$)}
  \label{fig:M_10}
\end{figure}

\begin{figure}[h]
  \centering
  \includegraphics[width=\textwidth,type=png,ext=.png,read=.png]{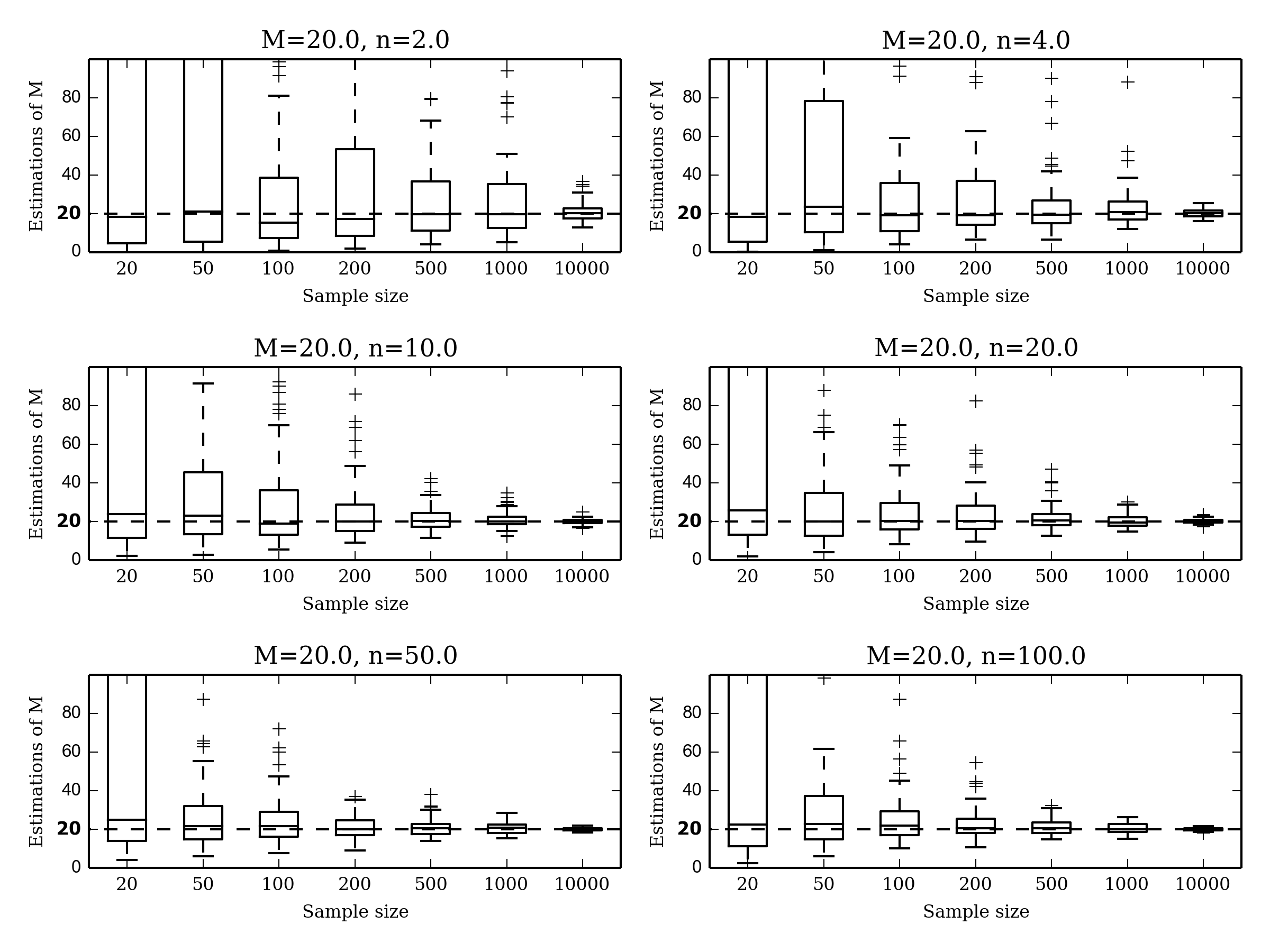}
  \caption{Estimations of $M$ (real value $M=20.0$)}
  \label{fig:M_20}
\end{figure}

\begin{figure}[h]
  \centering
  \includegraphics[width=\textwidth,type=png,ext=.png,read=.png]{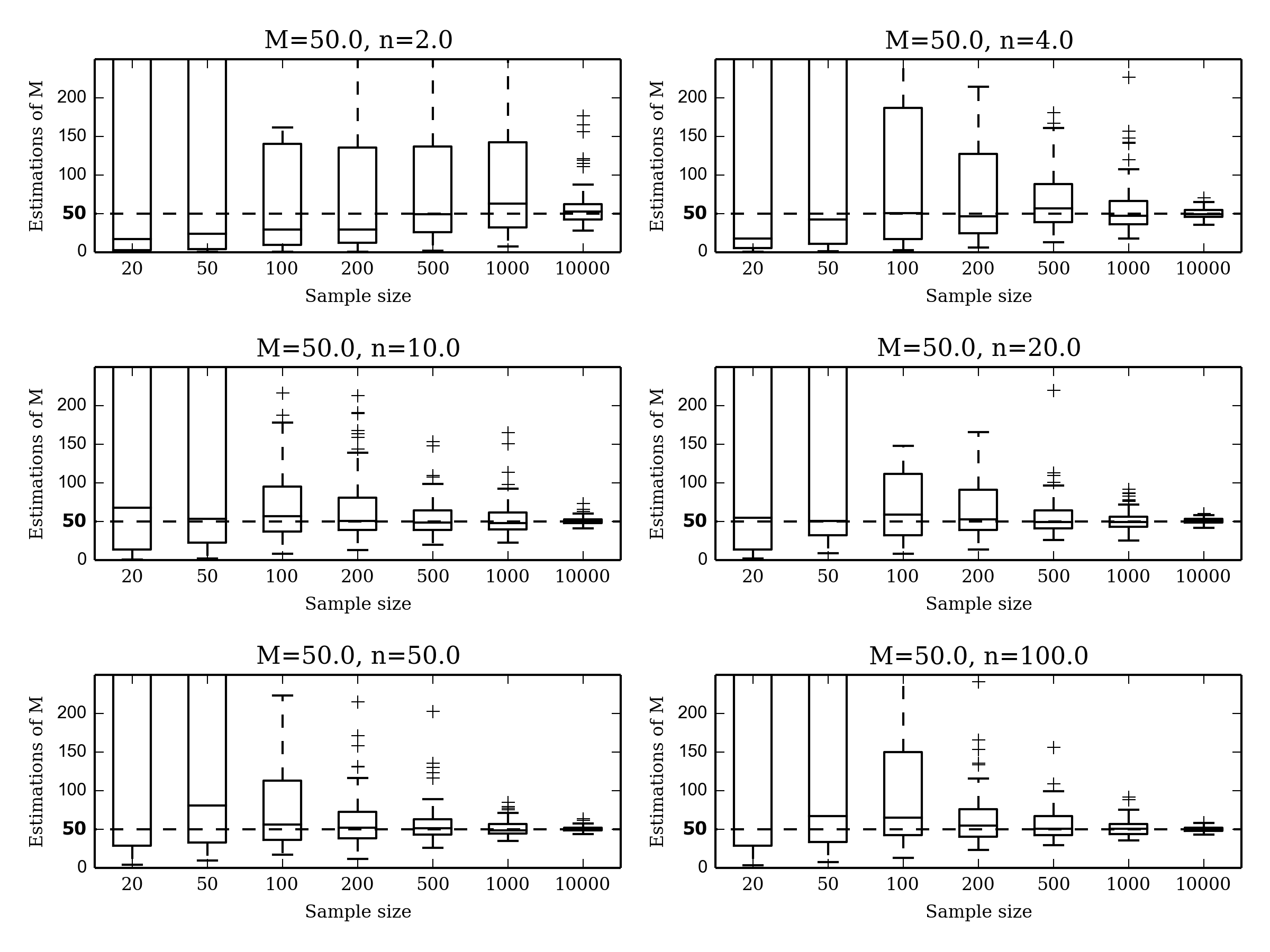}
  \caption{Estimations of $M$ (real value $M=50.0$)}
  \label{fig:M_50}
\end{figure}

\clearpage
Figures \ref{fig:M_0.1}, \ref{fig:M_0.2}, \ref{fig:M_0.5}, \ref{fig:M_1}, \ref{fig:M_5}, \ref{fig:M_10}, \ref{fig:M_20} and \ref{fig:M_50} show the results of the estimation of $M$ for all the values of $n$ and all the sample sizes, under the StSI model. The values where $M=(0.1, 0.2, 0.5, 1, 5, 10, 20, 50)$, $n=(2, 4, 10, 20, 50, 100)$ and $n_L=(20, 50, 100, 200, 500, 1000, 10000)$

\begin{figure}[h]
  \centering
  \includegraphics[width=\textwidth,type=png,ext=.png,read=.png]{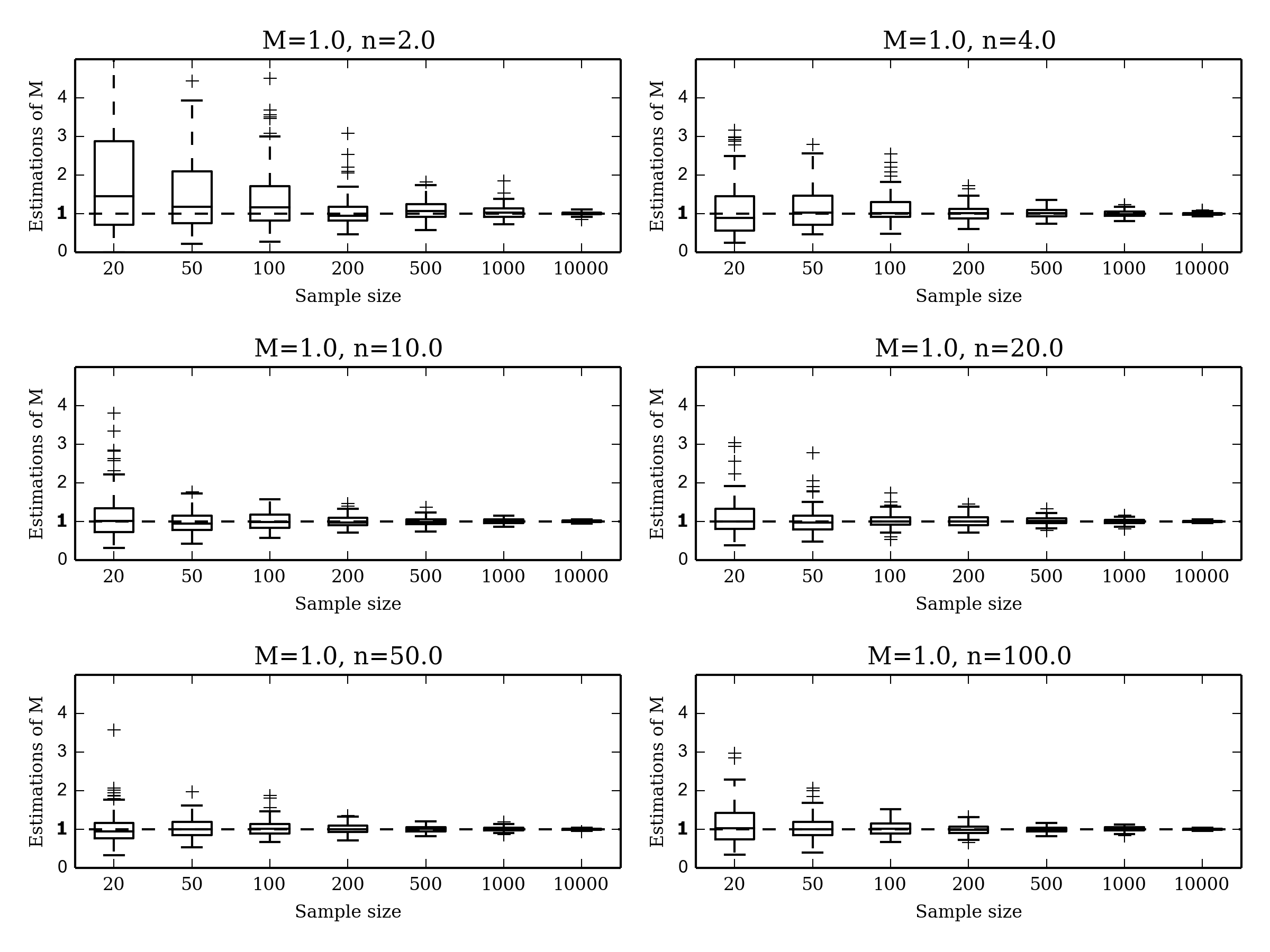}
  \caption{Estimations of $M$ (real value $M=2.0$)}
  \label{fig:M_2}
\end{figure}

\begin{figure}[h]
  \centering
  \includegraphics[width=\textwidth,type=png,ext=.png,read=.png]{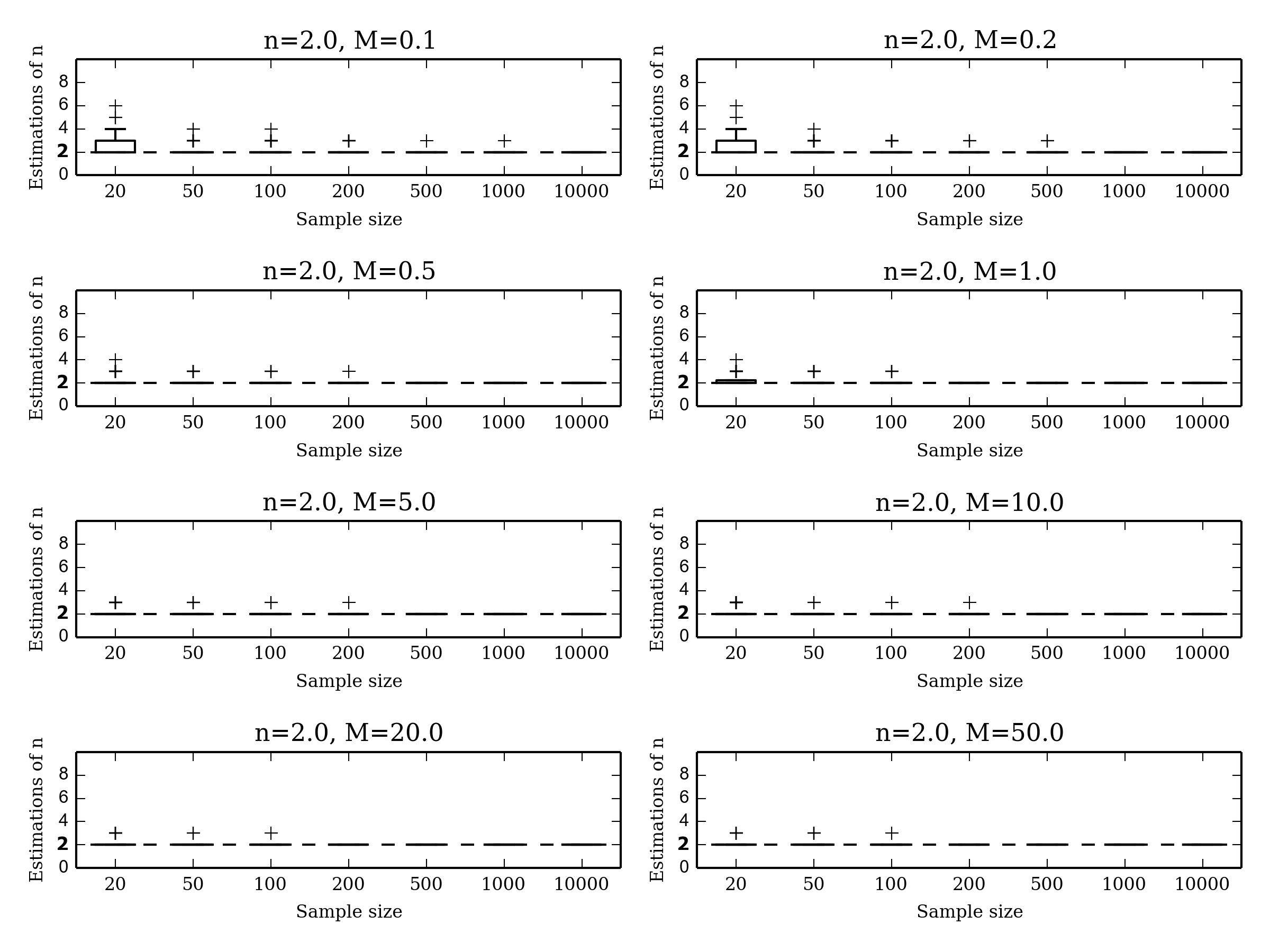}
  \caption{Estimations of $n$ (real value $n=2.0$)}
  \label{fig:n_2}
\end{figure}

\begin{figure}[h]
  \centering
  \includegraphics[width=\textwidth,type=png,ext=.png,read=.png]{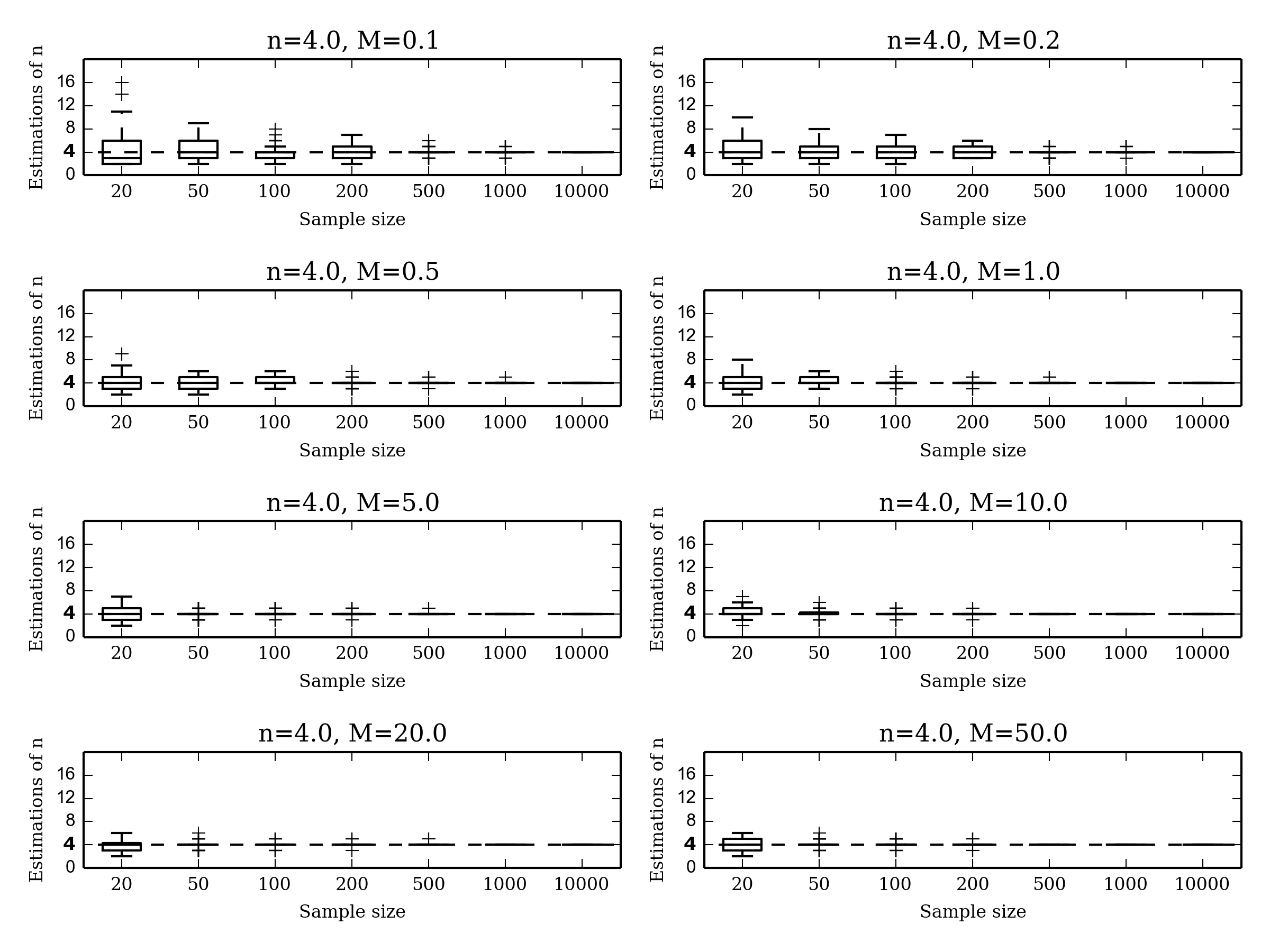}
  \caption{Estimations of $n$ (real value $n=4.0$)}
  \label{fig:n_4}
\end{figure}

\begin{figure}[h]
  \centering
  \includegraphics[width=\textwidth,type=png,ext=.png,read=.png]{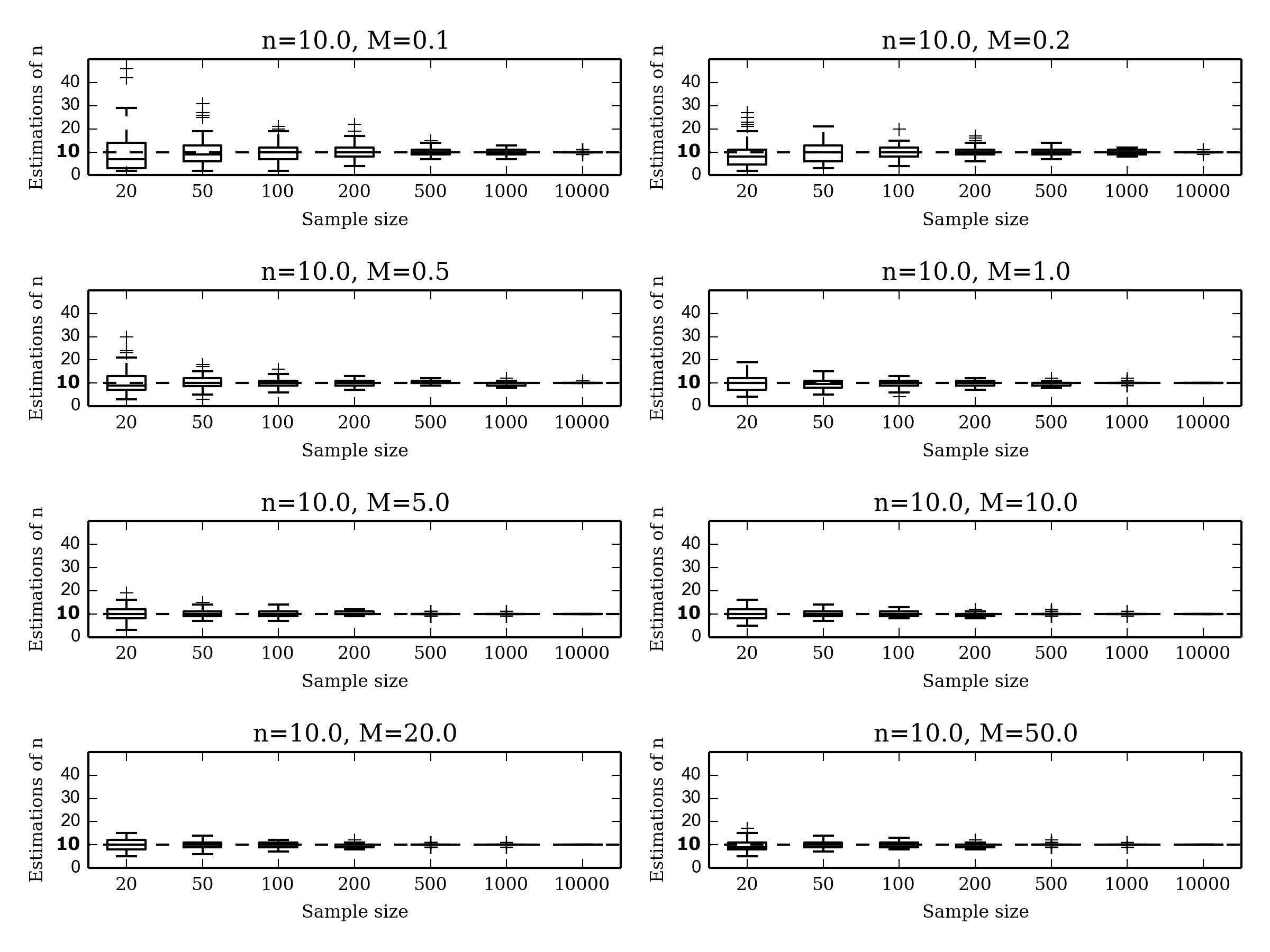}
  \caption{Estimations of $n$ (real value $n=10.0$)}
  \label{fig:n_10}
\end{figure}

\begin{figure}[h]
  \centering
  \includegraphics[width=\textwidth,type=png,ext=.png,read=.png]{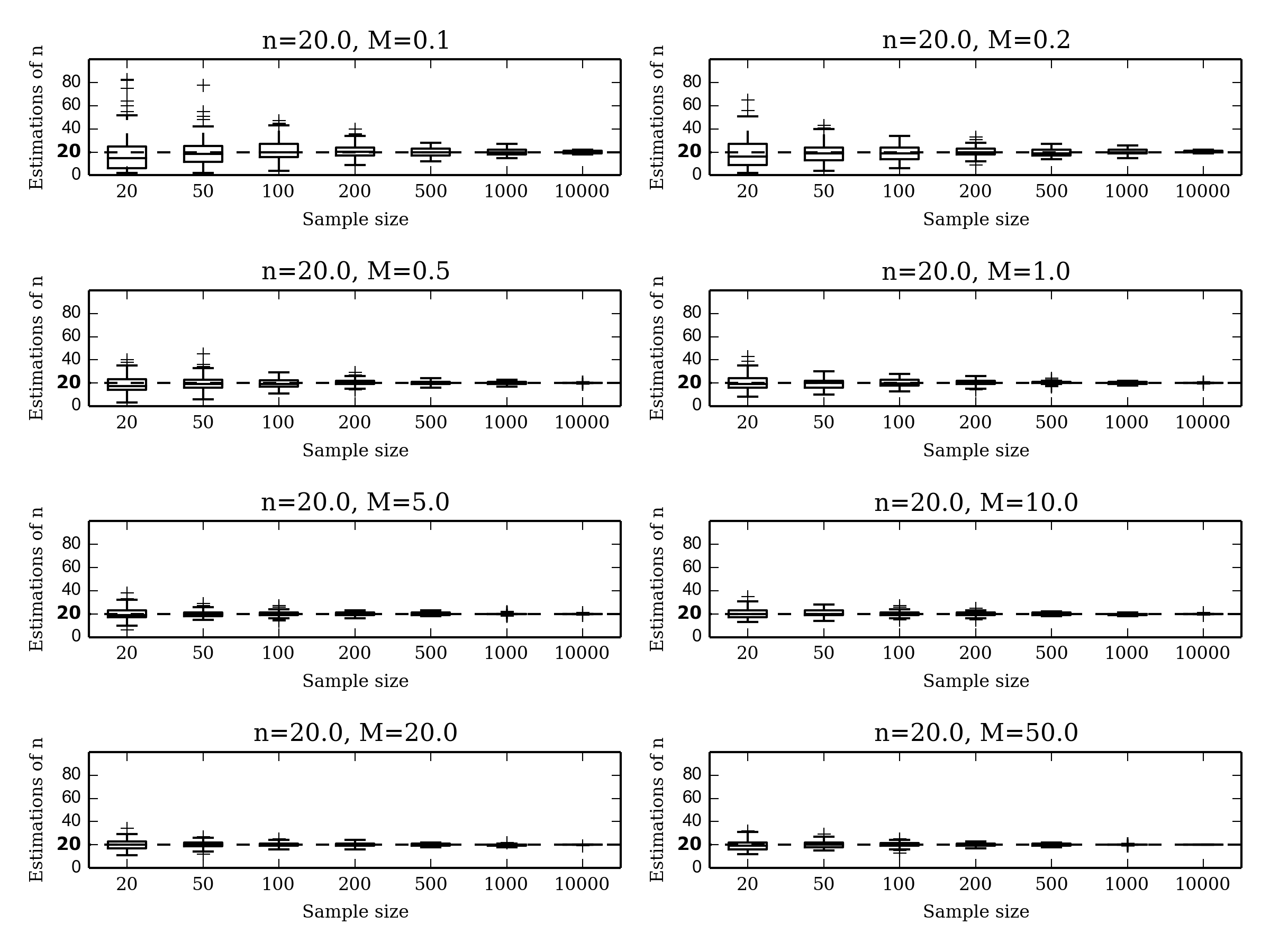}
  \caption{Estimations of $n$ (real value $n=20.0$)}
  \label{fig:n_20}
\end{figure}

\begin{figure}[h]
  \centering
  \includegraphics[width=\textwidth,type=png,ext=.png,read=.png]{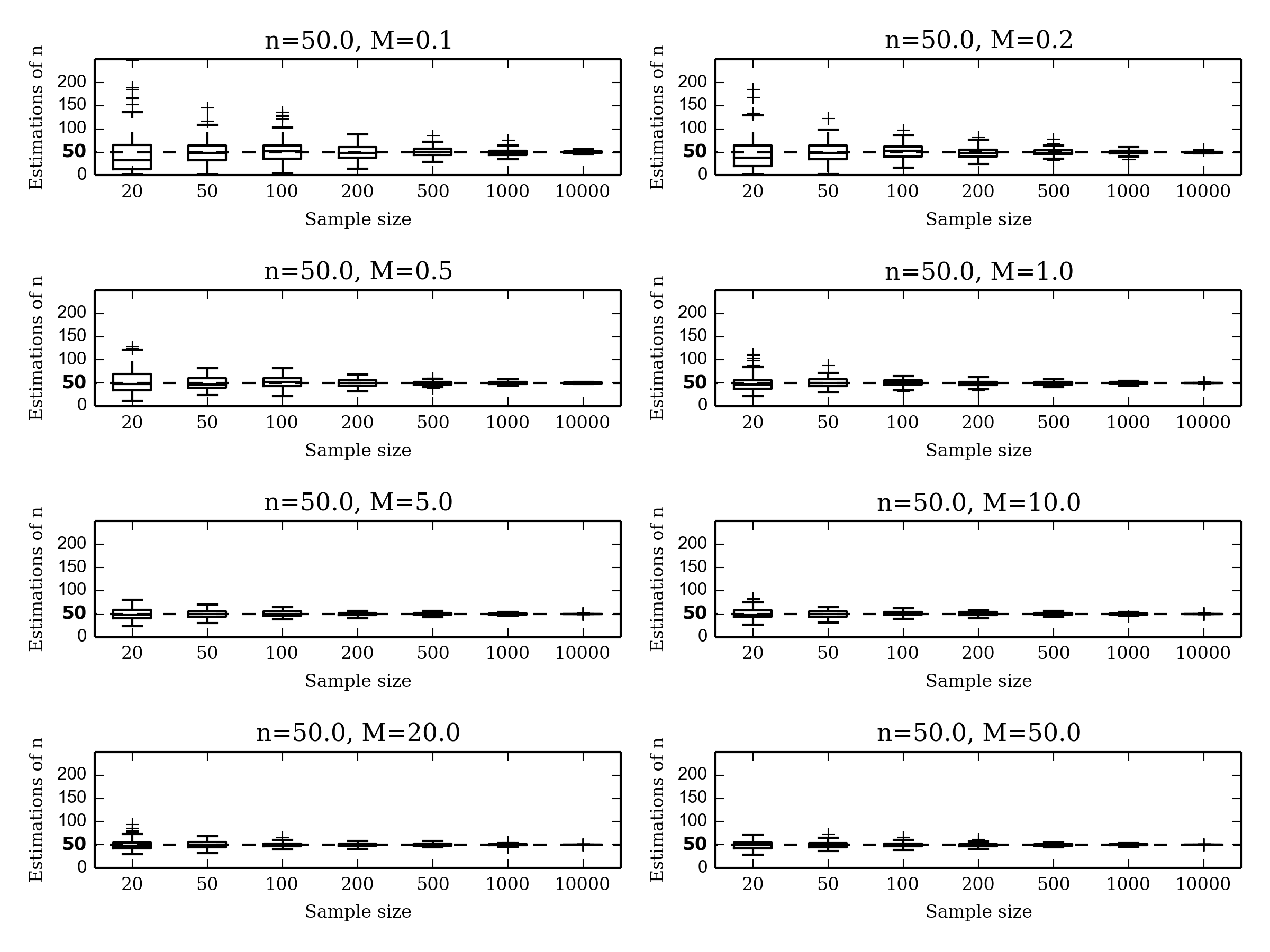}
  \caption{Estimations of $n$ (real value $n=50.0$)}
  \label{fig:n_50}
\end{figure}

\begin{figure}[h]
  \centering
  \includegraphics[width=\textwidth,type=png,ext=.png,read=.png]{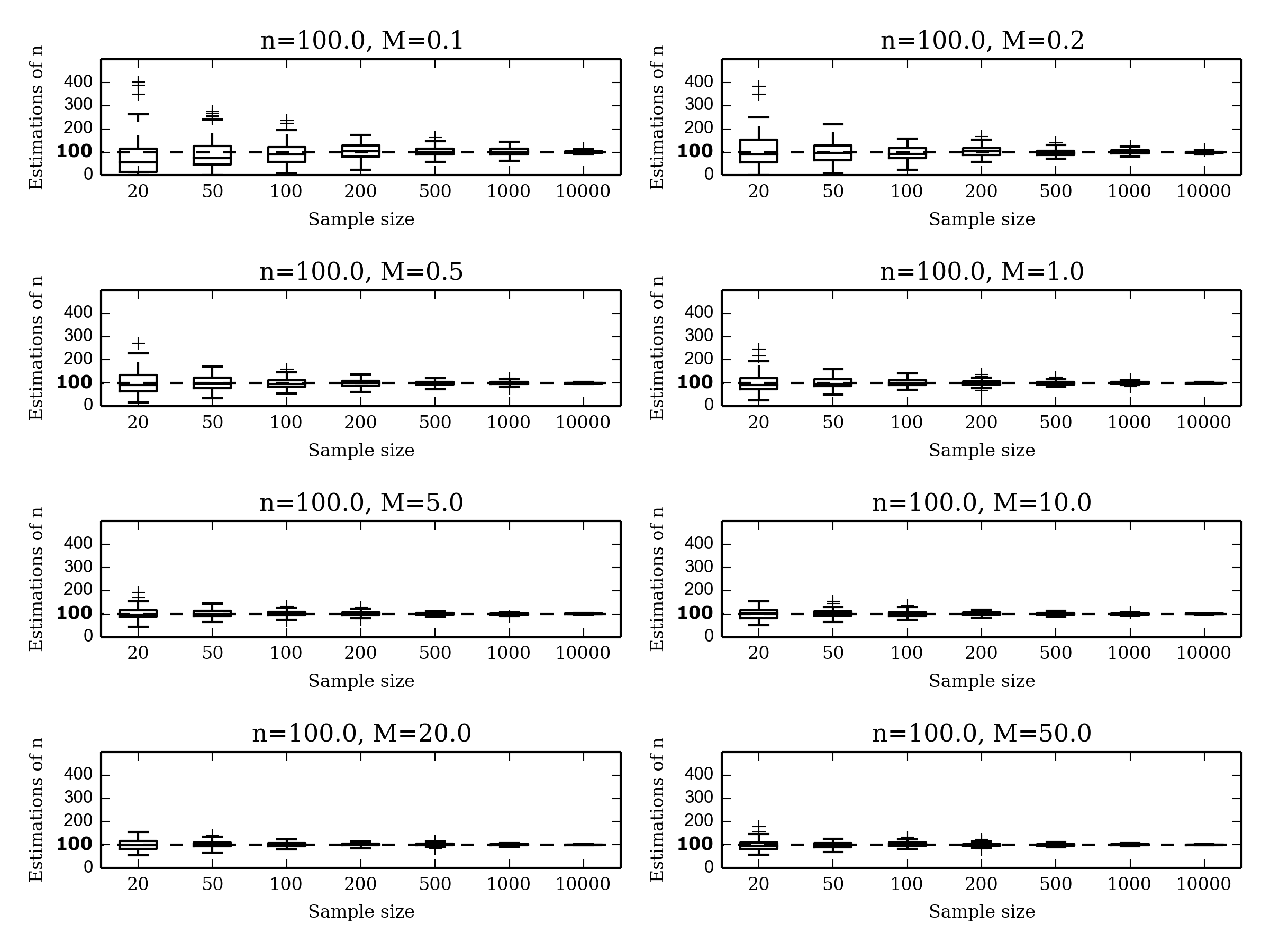}
  \caption{Estimations of $n$ (real value $n=100.0$)}
  \label{fig:n_100}
\end{figure}

\clearpage
Figures \ref{fig:n_2}, \ref{fig:n_4}, \ref{fig:n_10}, \ref{fig:n_20}, \ref{fig:n_50} and \ref{fig:n_100} show the results of the estimation of $n$ for all the values of $M$ and all the sample sizes, under the StSI model. The values where $M=(0.1, 0.2, 0.5, 1, 5, 10, 20, 50)$, $n=(2, 4, 10, 20, 50, 100)$ and $n_L=(20, 50, 100, 200, 500, 1000, 10000)$

